%% file: paper_appmine.tex
\documentclass[sigconf,twocolumn,nonacm,10pt]{acmart}

\renewcommand\footnotetextcopyrightpermission[1]{} 
\setcopyright{none}

\usepackage{graphicx}
\usepackage{multirow}

\usepackage{subcaption}
\newcommand{\arxiv}{}
\ifx\arxiv\undefined

\usepackage{times}
\usepackage{url}
\usepackage{epsfig}
\usepackage{xspace}

\usepackage{xcolor}
\usepackage{listings}
\usepackage{hyperref}
\fi

\newcommand{\ignore}[1]{}
\newcommand\sys{\ensuremath{\mathsf{AppMine}}}

\title{\sys: Behavioral Analytics for Web Application Vulnerability Detection}

\author{Indranil Jana and Alina Oprea}

\affiliation{%
  \institution{Khoury College of Computer Sciences, Northeastern University, Boston, MA, USA}
}

\begin{document}

\copyrightyear{2019}
\acmYear{2019}
\setcopyright{acmcopyright}
\acmConference[CCSW '19]{The ACM Cloud Computing Security Workshop}{November 11, 2019}{London, UK}
\acmBooktitle{2019 ACM Cloud Computing Security Workshop (CCSW 2019), November 11, 2019, London, UK}

\input{abstract}

\maketitle

\input{intro}

\input{background}
\input{methodology}

\input{experiments}
\input{related}

\input{conclusions}

\newpage
\bibliographystyle{plain}
\bibliography{references}
\input{appendix_short}

\end{document}

%% file: abstract.tex
\begin{abstract}

Web applications in widespread use have always been the target of large-scale attacks, leading to massive disruption of services and financial loss, as in the Equifax data breach. It has become common practice to deploy web application in containers like Docker for better portability and ease of deployment. We design a system called \sys\ for lightweight monitoring of web applications running in Docker containers and detection of unknown web vulnerabilities. \sys\ is an unsupervised learning system, trained only on legitimate workloads of web application, to detect anomalies based on either traditional models (PCA and one-class SVM), or more advanced neural-network architectures (LSTM). In our evaluation, we demonstrate that the neural network model outperforms more traditional methods on a range of web applications and recreated exploits. For instance, \sys\ achieves average AUC scores as high as 0.97 for the Apache Struts application (with the CVE-2017-5638 exploit used in the Equifax breach), while the AUC scores for PCA and one-class SVM are 0.81 and 0.83, respectively. 



\end{abstract} 

%% file: intro.tex
\section{Introduction}

Web-based attacks remain one of the major attack vectors with notorious security incidents such as the Equifax breach and Drupalgeddon 
being attributed to web application vulnerabilities~\cite{Struts}. Such attacks already resulted in serious breaches of confidential and personal information affecting consumers and businesses alike. For instance, the Equifax security incident from 2017 impacted approximately 143 million U.S. consumers, compromising their identity (e.g., SSN, driver license) and financial information (e.g., credit card numbers)~\cite{Equifax}. We thus expect that cyber security attacks will induce even more devastating consequences in the future.

The majority of web applications today are distributed in Docker containers \cite{Docker} to run in either public cloud or private cloud environments. While containers offer flexibility, scalability, and have low performance overhead, on the downside existing defenses for enterprise networks (e.g., anti-virus, intrusion prevention systems, web proxies) are not readily applicable in container-based environments. Existing solutions based on static analysis, dynamic analysis, input validation, and fuzz testing have well-known limitations.

In this paper we design an unsupervised learning framework called \sys\ for web application defense that is effective at protecting container-based applications against a range of vulnerabilities. Our framework monitors web applications deployed in Docker containers and collects detailed run-time information using well-known monitoring tools such as Sysdig~\cite{Sysdig}. At the core of the technical approach lies a machine learning framework that uses Recurrent Neural Network (RNN) architectures popular in the deep learning community for representing sequential dependencies in time-series data. We use a type of RNN called Long-Short Term Memory (LSTM), which can capture the time evolution of application profiles, learn both short-term and long-term dependencies, and identify security anomalies that deviate from historical behavior. 

 To evaluate our techniques, we deploy a testbed in which we set up four popular web application in Docker containers, and recreate seven exploits, using Metasploit modules. We collect system call data using the Sysdig monitoring agent. We compare our LSTM model with two traditional anomaly detection algorithms: Principal Component Analysis (PCA) and One-Class Support Vector Machines (OCSVM). We demonstrate the effectiveness of our framework at detecting well-known vulnerabilities such as those in the Apache Struts and Drupal applications with low false positive rates. Our LSTM model significantly improves upon more traditional anomaly detection models, by exploiting the dependencies in sequences of system calls over time. For instance, for the Apache Struts web application with the CVE-2017-5638 exploit used in the Equifax breach, the LSTM model achieves an AUC of 0.97, while PCA and OCSVM have AUCs of  0.81 and 0.83, respectively. The advantages of our \sys\ framework are that it does not require attack data for training, and it can be applied to detect unknown vulnerabilities in web applications.

  To summarize, our contributions are:
 \begin{enumerate}
 \item We design an unsupervised	 learning framework called \sys\ for detection of a range of web application vulnerabilities.
 \item We set up a testbed with four web applications and recreate seven exploits, collecting monitoring data from Sysdig agents deployed in Docker containers.
 \item We compare the performance of our neural-network LSTM model with that of traditional unsupervised learning models (PCA and OCSVM) and demonstrate its effectiveness. 
 \end{enumerate}

\paragraph{Organization} We provide background on web application vulnerabilities and our threat model in Section~\ref{sec:background}. We describe our methodology, testbed setup, data collection, and machine learning framework in Section~\ref{sec:method}. We evaluate our system \sys\ on a range of web applications and vulnerabilities in Section~\ref{sec:eval}. We present related work in Section~\ref{sec:related} and conclude in Section~\ref{sec:conclusions}.


%% file: background.tex
\section{Background and Overview}
\label{sec:background}

We first provide some background on web application vulnerabilities, then discuss our threat model, and give an overview of our system \sys.

\subsection{Web application vulnerabilities}

 Web applications can be deployed either on-premise (in private clouds and data centers) or off-premise (in public cloud environments). A recent trend is to \emph{containerize} web applications, which involves running them in Docker containers for increased portability across different hardware and software stacks.

Vulnerabilities in web and database applications might expose enterprise networks to serious security breaches. For instance, several remote code execution vulnerabilities (CVE-18-9805 and CVE-18-11775) have been discovered for Apache Struts, a Java  open source framework for developing web applications. Among these, the famous Apache Struts vulnerability CVE-2017-5638 allows remote attackers to execute arbitrary commands on the web server via the Content Type HTTP header. After getting access to the web server, attackers start propagating laterally in the network and reach the target of interest (usually a database storing confidential information). Other web vulnerabilities include SQL injection, cross-site scripting, and cross-site request forgery. 

Defenses against these application vulnerabilities usually involve patching vulnerable applications, after the exploit is known and a patch is available. Web application security testing tools use a combination of static code analysis~\cite{Pixy}, dynamic checks~\cite{XSSDynamic}, input validation~\cite{Scholte12}, and fuzz testing~\cite{Allister08}, but in general they have a number of well-known limitations (e.g., high false positives or false negatives). In general, despite all existing defenses, enterprises are still exposed against zero-day vulnerabilities in their web applications.



\subsection{Threat model}

We consider a system model in which web applications run in container environments such as Docker and monitoring agents are deployed in the containers. We assume that the containers themselves and the monitoring agents that collect the monitoring data are not under the attacker's control. Attackers are performing their actions remotely, interacting with the web application via network packets. We also assume that the adversary cannot tamper with the collected system call data. Attackers with access to the monitoring environment and the system logs are much more powerful, and are beyond our current scope. 

We thus handle scenarios in which attackers interact remotely with the web application, attempting to exploit a vulnerability, and get full access to the application running environment (i.e., the Docker container in our setup). The web application attack is usually an entry point for the attacker, interested in moving laterally in the environment (cloud or enterprise) and obtaining access to critical resources. For instance, in the Equifax breach, the Struts exploit allowed the attacker control of the web server, but the attacker's ultimate goal was obtaining access to the personal information of millions of customers.

\subsection{Overview}


\begin{figure*}[t]
\begin{centering}
\includegraphics[width=0.8\linewidth]{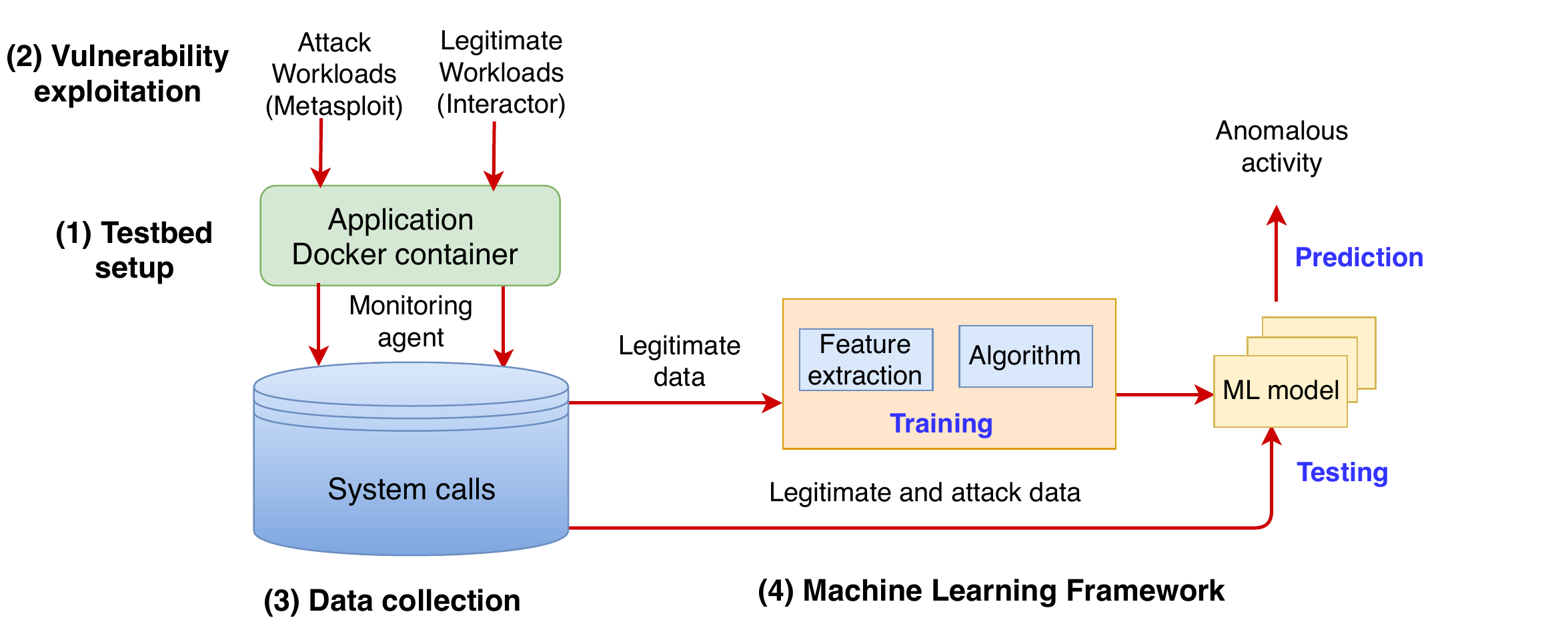}
\caption{Overview of the \sys\ system architecture.}
\label{fig:system}
\end{centering}
\end{figure*}

In designing our machine learning framework for web application threat detection, we leverage several main insights. First, there is a large amount of sequential dependence in application behavior manifested in unique, highly distinguishable sequences of system calls generated by an application. We argue that machine-learning techniques applied to application monitoring independently
of their temporal ordering, long-term dependencies, and contextual information do not
offer sufficient protection against advanced attacks. Therefore, we design machine learning architectures that leverage the sequential and contextual dependencies in application monitoring data. Second, we believe that application exploits and cyber attacks will result in deviation of monitoring data compared to an application running under normal conditions without exposure to the attack or vulnerability. To the extent the attack is observable in the collected monitoring data depends on a number of factors including the attack specifics and the granularity of collected data.  Third, by encapsulating single applications in each container, we can reduce the amount of noise from events generated by other applications, and create ``clean profiles'' of typical enterprise applications.

A number of challenges need to be addressed in designing our machine learning-based application threat detection framework. We highlight among them: (1) privacy and performance considerations that
prevent full-blown application monitoring; (2) dealing with non-deterministic application behavior; (3)  designing the most appropriate machine learning models for this setting. There is a lot of research and guidance on designing neural network architectures for other domains (e.g., image classification, speech recognition, machine translation), but there is no widely accepted methodology for security datasets that exhibit different semantics.

Figure~\ref{fig:system} provides an overview of our machine learning framework including the following components: (1) {\bf Testbed setup} to run various web applications and emulate legitimate behavior; (2) {\bf Vulnerability exploitation} to reproduce existing vulnerabilities using Metasploit; (3) {\bf Data collection} using the Sysdig monitoring agent installed in the Docker container environment; (4) {\bf Machine Learning Anomaly Detection Framework} supporting both traditional and deep-learning based models. We detail each of these components in the next section.


\ignore{
\begin{figure*}
\vspace{-0.5cm}
\begin{centering}
\includegraphics[width=0.8\textwidth]{figures/systemarch2.pdf} 
\caption{Overview of the \sys\ system architecture.}
\label{fig:system}
\vspace{-0.3cm}
\end{centering}
\end{figure*}
}

%% file: methodology.tex
\section{Methodology}
\label{sec:method}

In this section we discuss our system methodology. We start by describing the testbed setup, the vulnerability exploitation method, and the data collection. Finally, we describe in detail our unsupervised machine learning framework for web application vulnerability detection.

\subsection{Testbed setup}

Cloud providers that deploy containerized web applications might leverage existing monitoring tools and obtain data from realistic user interactions with the web applications. However, potential privacy implications prevent these providers to perform detailed monitoring and release their datasets to the broader community. As we do not have access to datasets collected from deployed web applications, we create our own monitoring environment for applications deployed in Docker containers.  We set up an Ubuntu 16.04 Virtual Machine running Docker and install the Sysdig monitoring agent. For each considered web application, we built and deployed Docker containers on the Ubuntu machine. We used a separate Kali Linux 2 based Virtual Machine to run the scripts for generating the attack data using Metasploit. We discuss now in more details the choice of web applications, how we generate realistic legitimate workloads, and how we set up the exploit emulation.


\paragraph{Web applications}

The web applications we selected include: Apache Struts~\cite{Struts} (two different variants), Drupal~\cite{Drupal},~WordPress\cite{WordPress} (three different plugins~\cite{wpajaxloadmore,wpnmedia,wpreflexgallery}), and ProFTPD~\cite{ProFTPD}. All of these are popular, open-source web applications that experienced vulnerabilities in the last few years. 

To select exploits of interest for these applications, we evaluated recent exploits from the CVE database and identified a set of seven exploits that had Metasploit modules. Table \ref{table:testedapps} details the application description and vulnerabilities that we emulate in our environment. We include CVE numbers for five out of the seven exploits, as the other two do not yet have a CVE number assigned. The CVE-2017-5638 for Apache Struts is the famous remote code execution vulnerability that led to the Equifax data breach. 
\\

\begin{table*}
\begin{center}
\begin{tabular}{ |c|c|c|c| }
 \hline
App Name & Description & Plugin Name & CVE\\
 \hline
\multirow{2}{*}{Struts} & \multirow{2}{*}{Framework for developing Java EE webapps} & \multirow{2}{*}{-} & CVE-2017-5638 \\
 \cline{4-4}  & & & CVE-2017-9805\\
\hline
Drupal &  Content-management framework & - & CVE-2018-7600 \\
 \hline
\multirow{3}{*}{WordPress} & \multirow{2}{*}{Content management system} & Reflex Gallery Plugin & CVE-2015-4133\\
\cline{3-4}& & Ajax Load More Plugin & -\\
\cline{3-4}& & N-Media Website Contact Form & -\\
\hline
ProFTPD & FTP server & - & CVE-2015-3306\\

 \hline
\end{tabular}
\caption{Open-source web applications and vulnerabilities considered for this work.}
\label{table:testedapps}
\end{center}
\end{table*}



\paragraph{Legitimate workloads}



One of the main challenges in our work is setting up workloads and interactions with the web applications that are similar to those generated by actual users. For this purpose, we found traditional web crawling to be unsuitable, as it focuses more on gathering data in an automatic manner than on creating realistic, human-like interaction. Thus, we designed {\bf Interactor}, a program that uses Selenium \cite{Selenium} to realistically interact with elements on the web interface of the web application. {\bf Interactor} opens a browser instance per user and interacts with the page by filling available forms, clicking hyperlinks, and pressing buttons in a randomized order. Small, random delays of 1-5 seconds are also inserted during the operation of {\bf Interactor}, so it simulates user behavior more realistically than a web scraper. Data collection for Struts, Drupal and Wordpress was done using {\bf Interactor}. For ProFTPD, we chose to leverage {\bf ftpbench} \cite{ftpbench} for data generation, a benchmark tool that enables user login to an FTP sever and  uploading files of different sizes.

\subsection{Vulnerability exploitation}


The Metasploit Framework \cite{Metasploit} is an open source project that serves as a penetration testing platform for finding and exploiting vulnerabilities.
For our work, we used existing Metasploit modules for the tested vulnerabilities and modified them as necessary to work on our versions of the web applications. Additionally, Metasploit comes with a set of post-exploitation information gathering modules that can be run on compromised machines, and we used them after exploiting each web application to ensure that we gained access to the victim. This step replicates a likely scenario in real-world exploitation, where information gathering is often the first step an attacker takes after compromising a system, so as to learn the victim's system and network configuration and discover more vulnerabilities. Table \ref{table:attackscripts} lists the post-exploitation scripts we tested and collected data for. They include gather network information, collect system information, user list, and credentials, and dump password hash files.

\begin{table*}
\begin{center}
\begin{tabular}{ |c|c| }
 \hline
Post-exploit script name & Description\\
 \hline
checkcontainer & Check whether target running inside container\\
\hline
ecryptfs\_creds & Collect all users' .ecrypts directories\\
\hline
enum\_configs & Collect configuration files for Apache, MySQL, etc.\\
\hline
enum\_network & Gather network information\\
\hline
enum\_protections & Find antivirus/IDS/firewalls etc\\
\hline
enum\_psk & Collect 802-11-Wireless-Security credentials\\
\hline
enum\_system & Gather system information (e.g., installed packages)\\
\hline
enum\_users\_history & Gather user list, bash history, vim history, etc.\\
\hline
enum\_xchat & Gather XChat's configuration files\\
\hline
env & Collect environment variables\\
\hline
gnome\_commander\_creds & Collect cleartext passwords from Gnome-commander\\
\hline
hashdump & Dump password hashes for all users\\
\hline
mount\_cifs\_creds & Obtain mount.cifs/mount.smbfs credentials from /etc/fstab\\
\hline
pptpd\_chap\_secrets & Collect PPTP VPN information\\
\hline
tor\_hiddenservices & Collect TOR Hidden Services hostnames and private keys\\
\hline
\end{tabular}
\caption{Metasploit post-exploitation scripts used for generating attack data.}
\label{table:attackscripts}
\end{center}
\end{table*}

\subsection{Data collection}


We use Sysdig \cite{Sysdig} to monitor the activity of the  web applications running in Docker containers. Sysdig has the ability to collect sequences of system calls made by the application in the container.

For each application except ProFTPD, we ran our {\bf Interactor} program to emulate 1, 3, 5, 10 and 15 simultaneous users. Each user performed a number of actions chosen randomly between 50 and 100, where an action is an activity such as clicking on a hyperlink, button on the web page, or filling a form. Monitoring was started after running the containers, and ended when the last user had performed his last action. For ProFTPD, we used the login and upload benchmarks from ftpbench \cite{ftpbench} for 30 mins to 1 hour each to generate legitimate data. For attack data, Sysdig monitoring and data collection was done similarly.

\begin{figure*}[t]
\begin{centering}
\includegraphics[width=\textwidth]{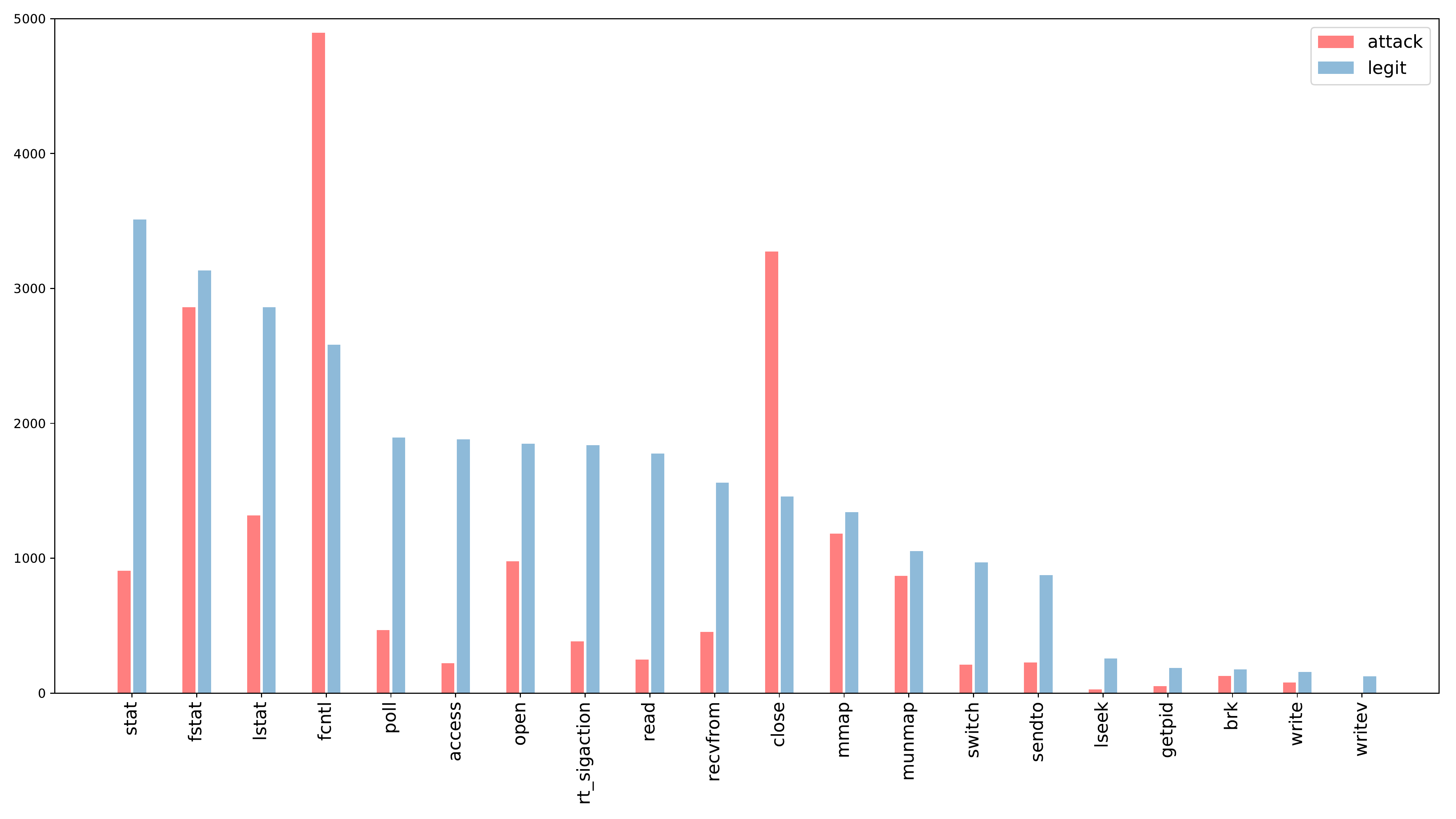}
\caption{System call distribution for ~15 seconds of data for WordPress, attack being shown in red is the enum\_network script}
\label{fig:highrocdist}
\end{centering}
\end{figure*}

Table \ref{table:traintestdatasizes} shows the data statistics for the four applications we monitored. In total, we collected between 300 and 360 minutes of data for each applications and vulnerability.  We split the sessions into training and testing for the machine learning framework, to minimize correlation and ensure independence of training and testing data. In Figure \ref{fig:highrocdist}, we show the distribution of the top 20 system calls during legitimate use for the WordPress application over a duration of approximately 15 seconds. The attack is being performed against the N-media contact form plugin, and the post-exploit script being run is enum\_network. Clearly, the system calls during the attack script are noticeably distinguishable from the ones used during the regular application runs. Notably, some system calls (e.g., fcntl, close) are used more frequently during the attack. 

\begin{table*}
\begin{center}
\begin{tabular}{ |c|c|c| }
 \hline
App/Plugin Name & Training data & Legitimate testing data\\
 \hline
Struts CVE-2017-5638 & 282.42 mins & 13.46 mins \\
\hline
 Struts CVE-2017-9805 & 308.78 mins & 28 mins\\
\hline
Drupal CVE-2018-7600 & 301.76 mins & 17.45 mins\\
 \hline
WP Reflex Gallery Plugin & 323.15 mins & 16.54 mins\\
\hline
WP Ajax Load More Plugin & 317.27 mins & 22.42 mins\\
\hline
WP N-Media Website Contact Form & 321.05 mins & 18.64 mins\\
\hline
ProFTPD CVE-2015-3306 & 342.85 mins & 20 mins\\

 \hline
\end{tabular}
\caption{Amount of data used for training and validation.}
\label{table:traintestdatasizes}
\end{center}
\end{table*}

\subsection{Machine Learning Framework}
\label{sec:ml}


\sys\ uses an unsupervised learning framework for web vulnerability detection.
Our motivation in adopting unsupervised learning includes: (1) not relying on attack data in the training phase; and (2) the ability to generalize for detecting new, unseen vulnerabilities. However, it is well known that anomaly detection techniques in security typically generate large amounts of false positives and are challenging to tune in operational settings~\cite{sommer2010outside}. We address this challenge by using the sequential and temporal dependencies in application system call data.

\paragraph{Training and testing the models.} As shown in Figure~\ref{fig:system}, \sys\ leverages only legitimate application data for training a machine learning model that learns the system call distribution under normal conditions. As already mentioned, we vary the number of users and their actions to create a set of diverse legitimate workloads for training the models. \sys\ creates system-call based feature representations for time windows of fixed length. Depending on the ML technique, the feature vector for either one time interval or a sequence of time intervals are given as input to the ML algorithm. The output of the training phase is a model that can determine the likelihood of a certain sequence of system calls. Importantly, as each application exhibits different behavior, we train a model per application to learn the application normal behavior. 

\begin{figure*}[t]
\begin{centering}
\includegraphics[width=6in]{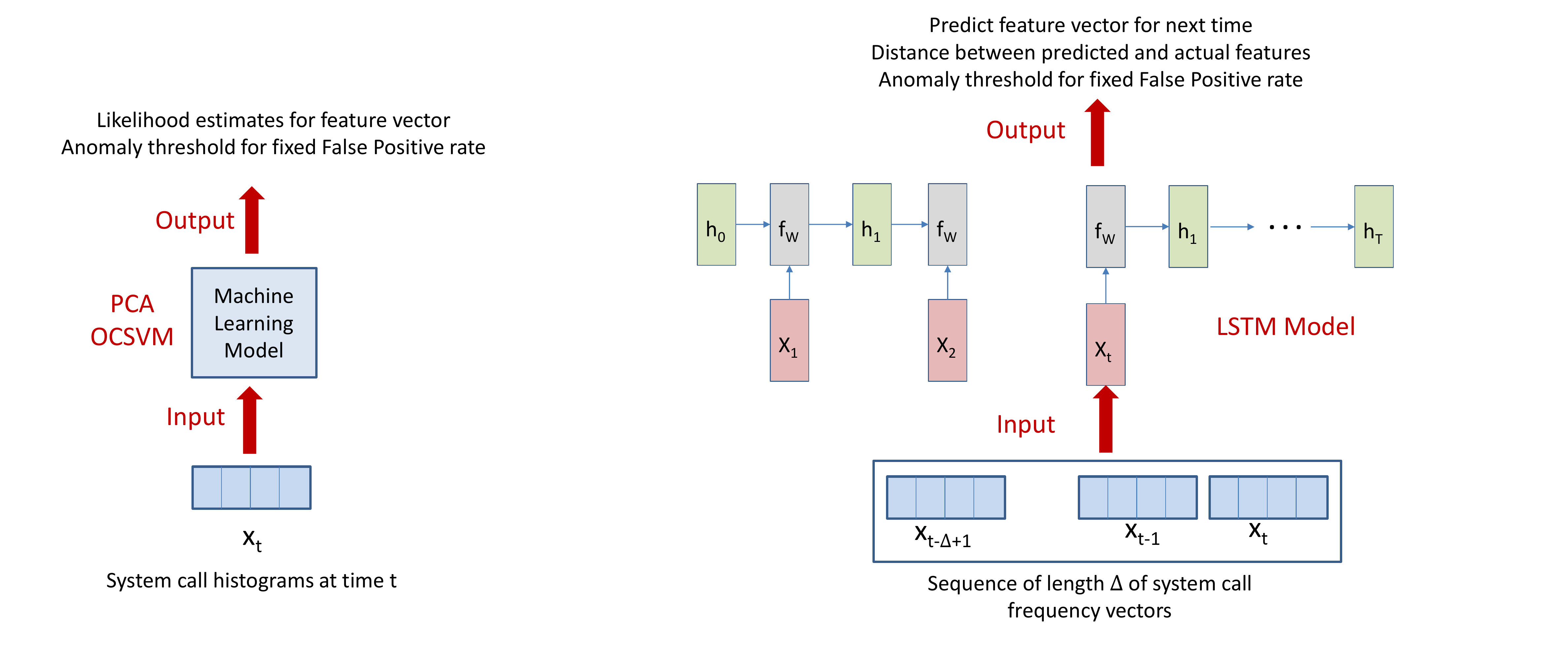}
\caption{Machine Learning methods in \sys. Traditional models on the left and the LSTM model on the right.}
\label{fig:ML}
\end{centering}
\end{figure*}

	At testing time, the application model is applied to new data generated by the same web application. We run the model on both legitimate and attack data for that application. The model produces an \emph{anomaly score} for the feature vectors at a particular time window, indicating the likelihood that the application has been exploited in that time interval.  Based on the labeled ground truth, we compute standard metrics such as True Positives, False Positives, and Area Under the Curve (AUC). A graphical representation of our anomaly detection models is given in Figure~\ref{fig:ML}.

\paragraph{Traditional anomaly detection techniques}

As mentioned, Sysdig collects time-stamped sequences of system calls for web applications. The first question we had to answer was how to represent this data for use in an anomaly detection system. A simple and fairly common representation (see, for example, \cite{dymshits2017process}) is to create a feature vector $x^t = [f_1^t,\dots,f_s^t]$ storing the frequency for each system call during a fixed time interval $t$. Here $s$ is the total number of system calls and $f_i^t$ denotes the number of times the $i$-th system call has been used during time window $t$. The interval length is a hyper-parameter of the system. System calls that did not appear in that interval had a frequency value of 0.  For intervals where no system calls were captured by Sysdig, a feature vector of all zeros was used. For example, let us assume that a program makes the following system calls during a one-second interval:


\begin{center}\texttt{futex,futex,open,write,close,open,read,close}\end{center}
Then, the feature vector for that interval is (system call \texttt{exec} is not used and has value 0)\\
\begin{center}
\begin{tabular}{c|c|c|c|c|c|}
\texttt{close} & \texttt{futex} & \texttt{open} & \texttt{write} & \texttt{read} & \texttt{exec}\\
\hline
2 & 2 & 2 & 1 & 1 & 0 \\
\end{tabular}
\end{center}


We use two standard anomaly detection techniques: Principal Component Analysis (PCA) and One Class Support Vector Machine (OCSVM) based on this frequency feature representation. Both of these methods take as input the system call frequency feature vector computed for one time interval. We use these methods to create baselines for comparison with our neural-network based anomaly detection model that leverages the sequence dependencies among system calls.\\

\noindent {\bf PCA:} Principal Component Analysis (PCA) is a technique that projects a dataset onto a set of \emph{principal components}, determined to maximize the variance observed in the training data. If the original data is given by matrix $X$ (with $n$ rows and $d$ columns, where $n$ is the number of data records in $X$ and $d$ is the size of the feature space), then PCA determines the top $k$ eigenvectors of the co-variance matrix $\Sigma = X^T X$. Let the matrix of the top $k$ eigenvectors be $W_k$ (storing one eigenvector per column). Then the projection of a point $x$ onto the space generated by the top $k$ principal components is given by $\hat{x} = W_k^T x$. The principal components have the property of minimizing the total reconstruction error. PCA can also be trained as a density estimation model, which estimates a probability distribution $p(x)$ over $x \in X$ (in our case for frequency feature vectors), based on Maximum Likelihood estimation~\cite{Bishop}. We use this density-estimation variant of PCA. In training we compute Gaussian estimates of the probabilities of system call frequency vectors, while in testing we compute log likelihoods for both attack and legitimate data. If a point has very low log likelihood in testing, then we consider it as an attack point. We vary the threshold for log likelihood in order to obtain ROC curves that evaluate True Positives as a function of fixed False Positive Rates.

\noindent {\bf OCSVM:} A Support Vector Machine (SVM) is a supervised learning model that is trained on data points from two different classes, in an attempt to find a separating hyperplane that separates the classes by the maximal margin. One-class SVM is trained on data from only one class (in our case, the legitimate class) and is used to detect novelties (or anomalies) based on the learned representation. We learn a one-class SVM model based on the legitimate training data, and use it to predict scores for testing data points (both attack and legitimate). In this setting, the scores are the distances to the separating decision boundary. Again, we vary the scores for which we predict attacks, to compute True Positives at fixed False Positive Rates.



\paragraph{Learning application profiles using LSTM}

To capture sequential dependencies in application monitoring data, we leverage Recurrent Neural Network (RNN) architectures. RNNs are deep learning models designed to map sequential and time series data to sequential outputs~\cite{RNN}. RNNs have been successfully applied for tasks such as machine translation, but it was observed that in simple RNN architectures several problems arise during training. In particular, vanishing gradients and exploding gradients might prevent the network from training accurately when gradients need to be back-propagated through multi-layer networks. Long Short-Term Memory (LSTM) networks~\cite{LSTM} are special types of RNNs that introduce special forget
gates to better preserve long-term dependencies by mitigating the vanishing gradient
problem. They maintain a hidden state $h_t$ at time $t$ that is updated using the current input: $(h_t,x'_{t+1})=f_W(h_{t-1},x_t)$. Here $W$ are the model weights, $f_W$ is the function applied inside the neural network (including the linear operation and activation function), $x_t$ is the input to the network at time $t$, and $x'_{t+1}$ is the prediction of feature vector at time $t+1$. 



We designed an LSTM architecture that learns a model $f_W$ mapping \emph{sequences of system call frequency vectors} to predictions of the next feature vector. In particular, given a sequence of size $\Delta$, the model takes as input the system call frequency vectors for the last $\Delta$ time windows: $x_t,x_{t-1},\dots,x_{t-\Delta+1}$ and predicts the next frequency vector: $x'_{t+1} \leftarrow f_W(x_t,\dots,x_{t-\Delta+1})$. During training, we have information about the true value of the system call frequency vector $x_{t+1}$ at time $t+1$ and we measure the distance between the predicted value and the actual value. We found that using a standard distance metric (such as Euclidean distance) $|| x_{t+1} - x'_{t+1} ||_2$ is not effective. Instead, we leverage ideas from Term Frequency - Inverse Document Frequency (TF-IDF) in information retrieval and we weight each system call by its inverse frequency when defining a new distance between the predicted and actual values: $||x'_{t+1}-x_{t+1}||_S = \sqrt{\sum_{i=1}^d \bar{f}_i(x^{t+1}_i-x'^{t+1}_i)^2}$, where $d$ is the number of system calls, and  $\bar{f}_i$ is the inverse-frequency of the $i^{th}$ system call as seen during training.





During training we run the LSTM model only on legitimate data $x_1,\dots,x_t,\dots$ (sequence of system call frequency vectors) and we compute all the weighted Euclidian distances between predicted and actual frequency vectors. We determine an anomaly threshold based on the distribution of distances for each application. The threshold is chosen to minimize the false positive rate (i.e., legitimate values $x_t$ with distance higher than the threshold $T$). This is equivalent to picking a threshold $T$ such as the probability $P[||x'_t-x_t||_S>T ] <= p$, for some fixed false positive rate $p$.

During testing, we run the LSTM model on new data $y_1,\dots,y_t,\dots$ (including legitimate and attack vectors), and predict at each time interval the frequency vector $y'_t = f_W(y_{t-1},\dots,F_{t-\Delta})$ based on the previous $\Delta$ frequency vectors: $y_{t-1},\dots,y_{t-\Delta}$. We consider the frequency vector at time $t$ an anomaly if the distance is higher than the threshold: $||y'_t - y_t||_S > T$.

%% file: experiments.tex
\section{Experimental Evaluation}
\label{sec:eval}

In this section, we evaluate our three anomaly detection models, using metrics such as True Positives at fixed False Positive rates, and Area Under the Curve (AUC). We first discuss hyper-parameter choice, then compare the performance of LSTM with that of the traditional ML models (PCA and OCSVM), and finally we present detailed attack detection results per attack script.

\paragraph{Configuring ML models}



We describe here some of the configuration options and hyper-parameter selection for our models. 

\begin{figure}[thb]
\includegraphics[width=0.8\linewidth]{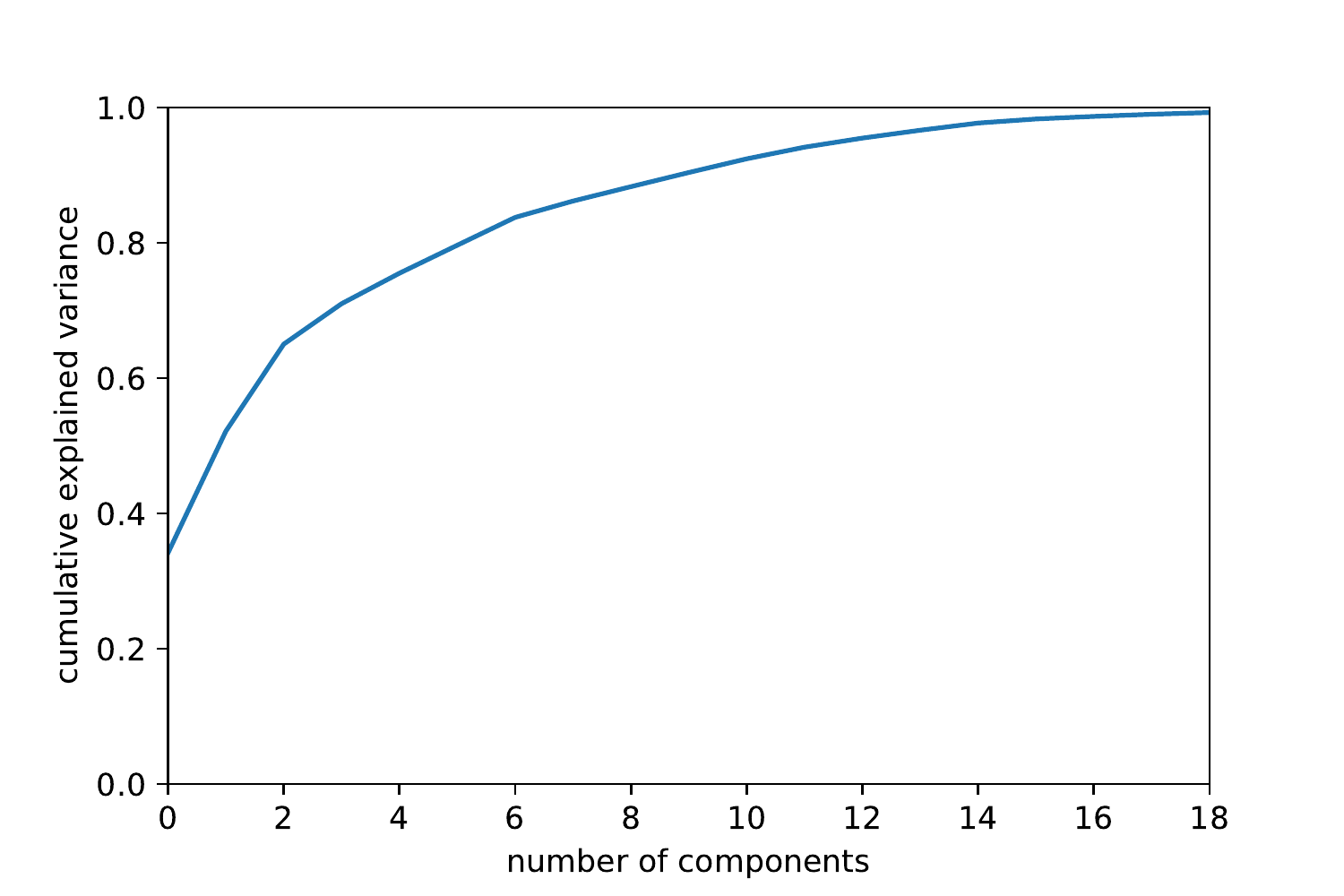}
\caption{Cumulative explained variance for PCA for Struts with CVE-2017-5638.}
\label{fig:PCAexplainedvariance}
\end{figure}

For PCA we need to select the number of components.
For all of the tested web applications, the cumulative variance graph started to flatten out at around 20 principal components. Figure \ref{fig:PCAexplainedvariance} shows an example for Struts (other applications being similar). Thus, we choose to represent the data using 20 principal components when running the PCA model. PCA assigns scores to each data point based on how far the projection of the data point on the lower dimensional space lies from the principal components. Normal, benign data points are expected to have a low score, while anomalous points will have higher scores.

As described in~\ref{sec:ml}, we trained our models on frequency count vectors. We tested time interval lengths of 100ms, 500ms, 1s, and 2s for creating feature vectors, and choose to work with one second as for the other options the results were worse (the ROC curves were closer to the diagonal).


\begin{figure}[th]
\centering
\includegraphics[width=0.8\linewidth]{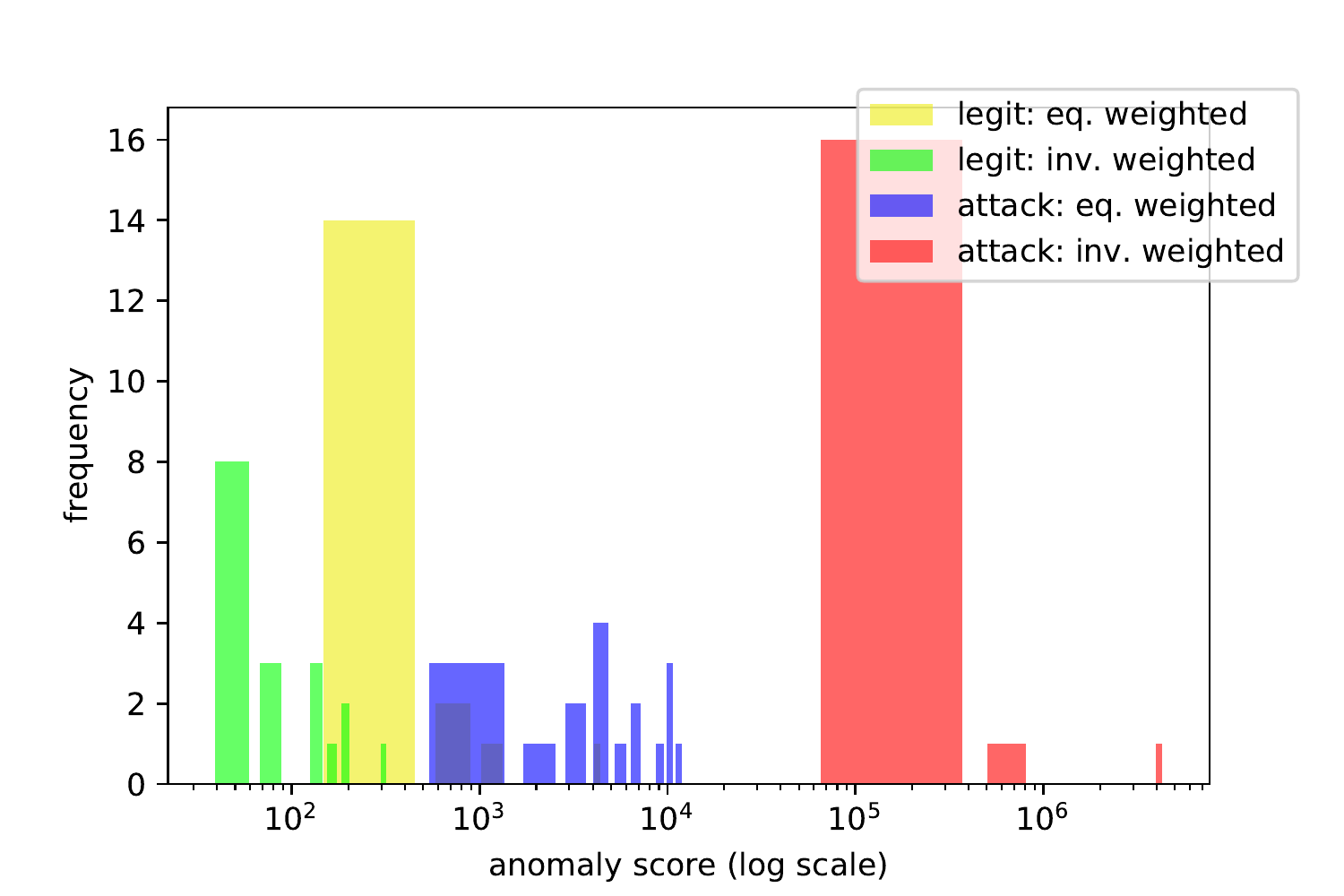}
\caption{Distribution of distances between predicted and actual frequency values for the {\sf enum\_configs} attack script for Struts with CVE-2017-5638.}
\label{fig:struts5638_scoredist}
\end{figure}

\begin{figure}[hb]
\centering
\includegraphics[width=0.8\linewidth]{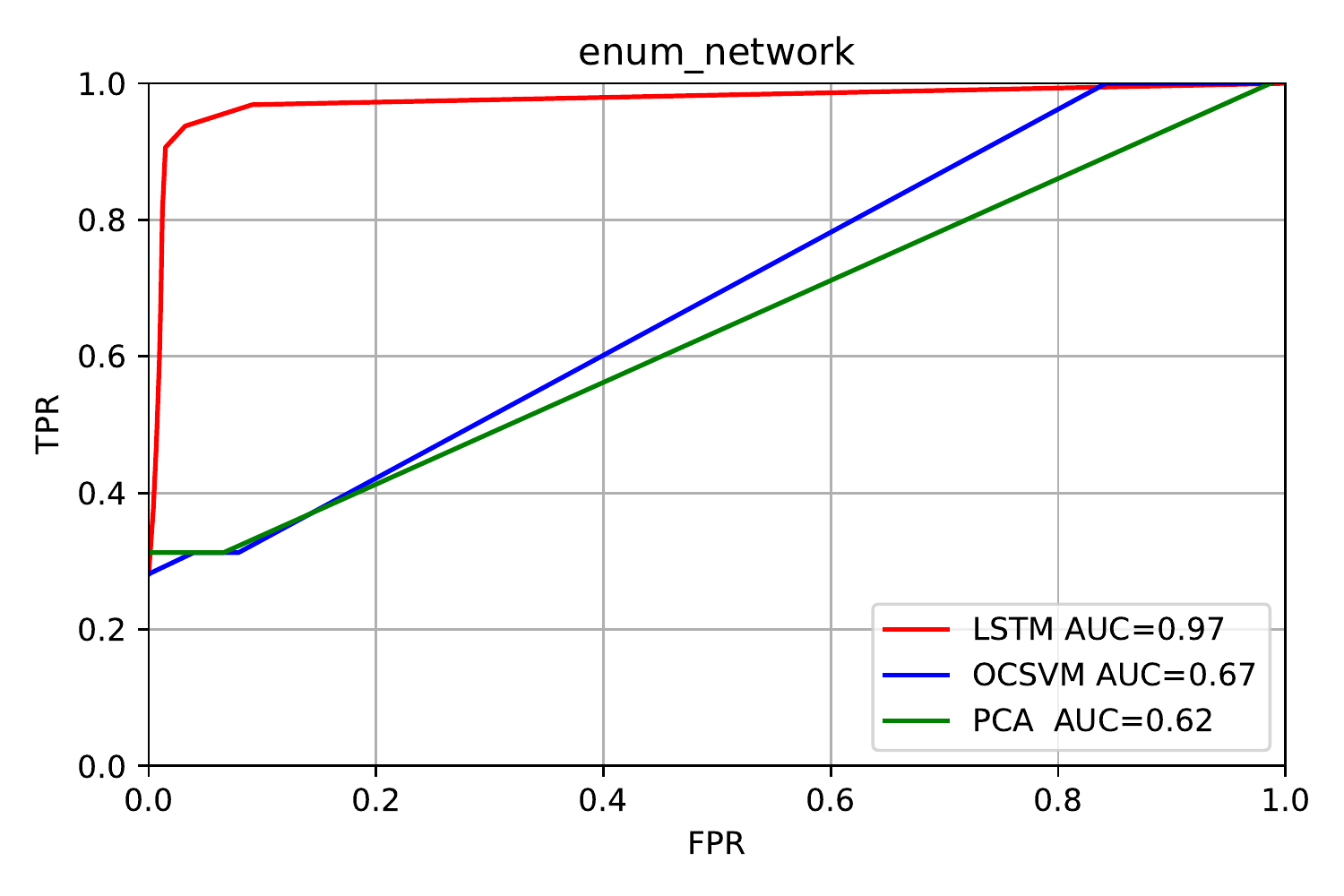}
\caption{ROC curves for $\mathsf{enum\_system}$ for Struts with CVE-2017-5638.}
\label{fig:strutsenumsystem}
\end{figure}

For training the LSTM models we used Python with Keras and Tensorflow to build a Sequential model with one LSTM layer with 100 hidden neurons and one Dense layer. We used Mean Squared Error (MSE) as the loss function with the ADAM optimization algorithm. Batch size of 128 was used while training over 150 epochs, with a 20\% validation split. Our hyper-parameters are given in  Table~\ref{table:hyperparams}. We used 15 seconds of history for training the LSTM model, implemented as 15 frequency vectors of length one-second each.



\begin{table}
\begin{center}
\begin{tabular}{ |c|c| }
 \hline
Parameter & Values\\
\hline
Number of hidden layers & 1 \\
Number of neurons & 100 \\
Batch size & 128 \\
Number of epochs & 150 \\
Timing window & 0.1s, 0.5s, 1s, 2s \\
Sequence size & 15 \\
\hline
\end{tabular}
\caption{Hyper-parameters for LSTM model.}
\label{table:hyperparams}
\end{center}
\end{table}

\begin{figure*}[t]
\centering
\begin{subfigure}{0.3\textwidth}
	\centering
	\includegraphics[width=\linewidth]{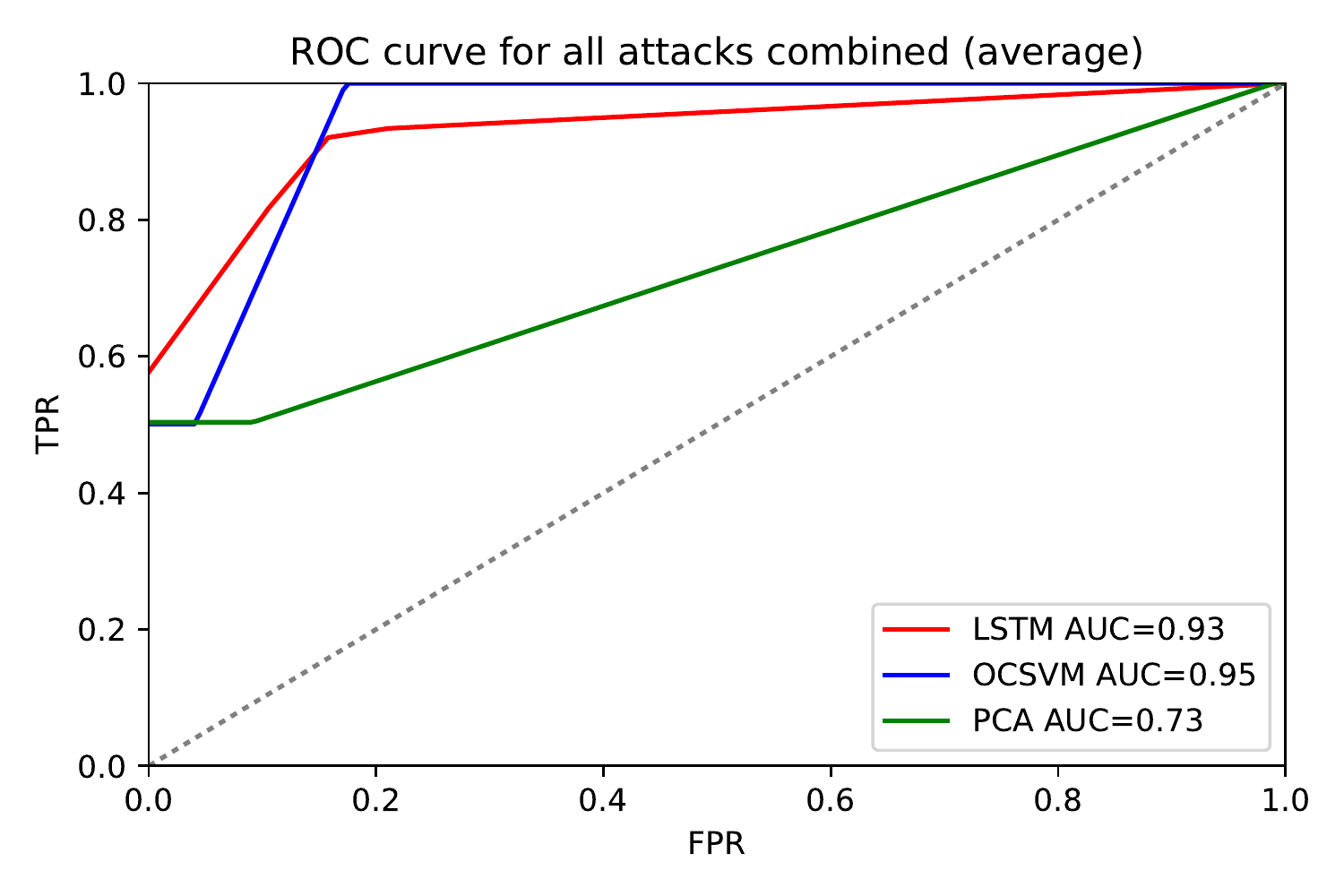}
	\caption{Drupal}
	\label{fig:drupal_allrocs}
\end{subfigure}
\hfill
\begin{subfigure}{0.3\textwidth}
         \centering
         \includegraphics[width=\linewidth]{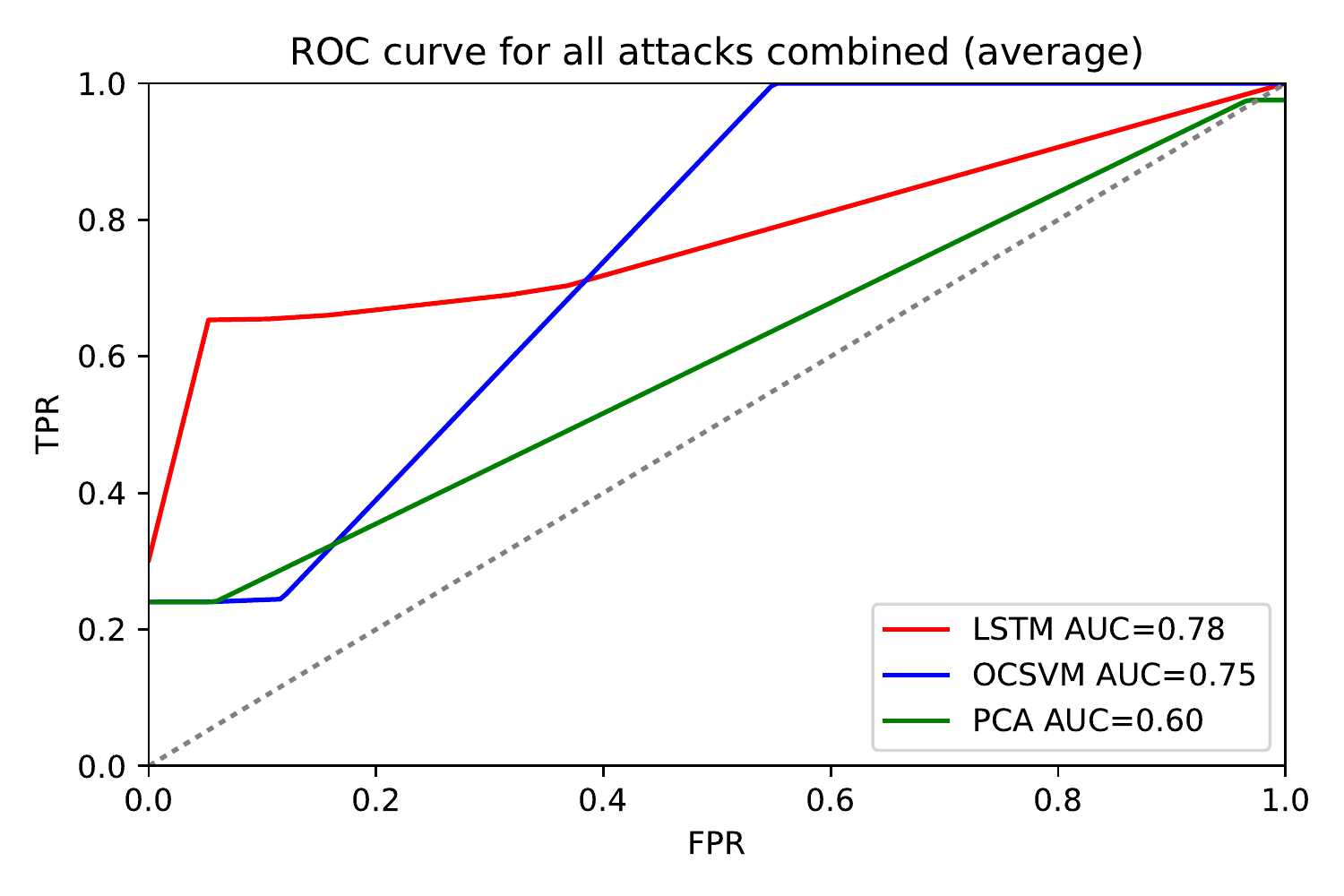}
         \caption{ProFTPD}
         \label{fig:proftpd_allrocs}
\end{subfigure}
\hfill
\begin{subfigure}{0.3\textwidth}
        \centering
         \includegraphics[width=\linewidth]{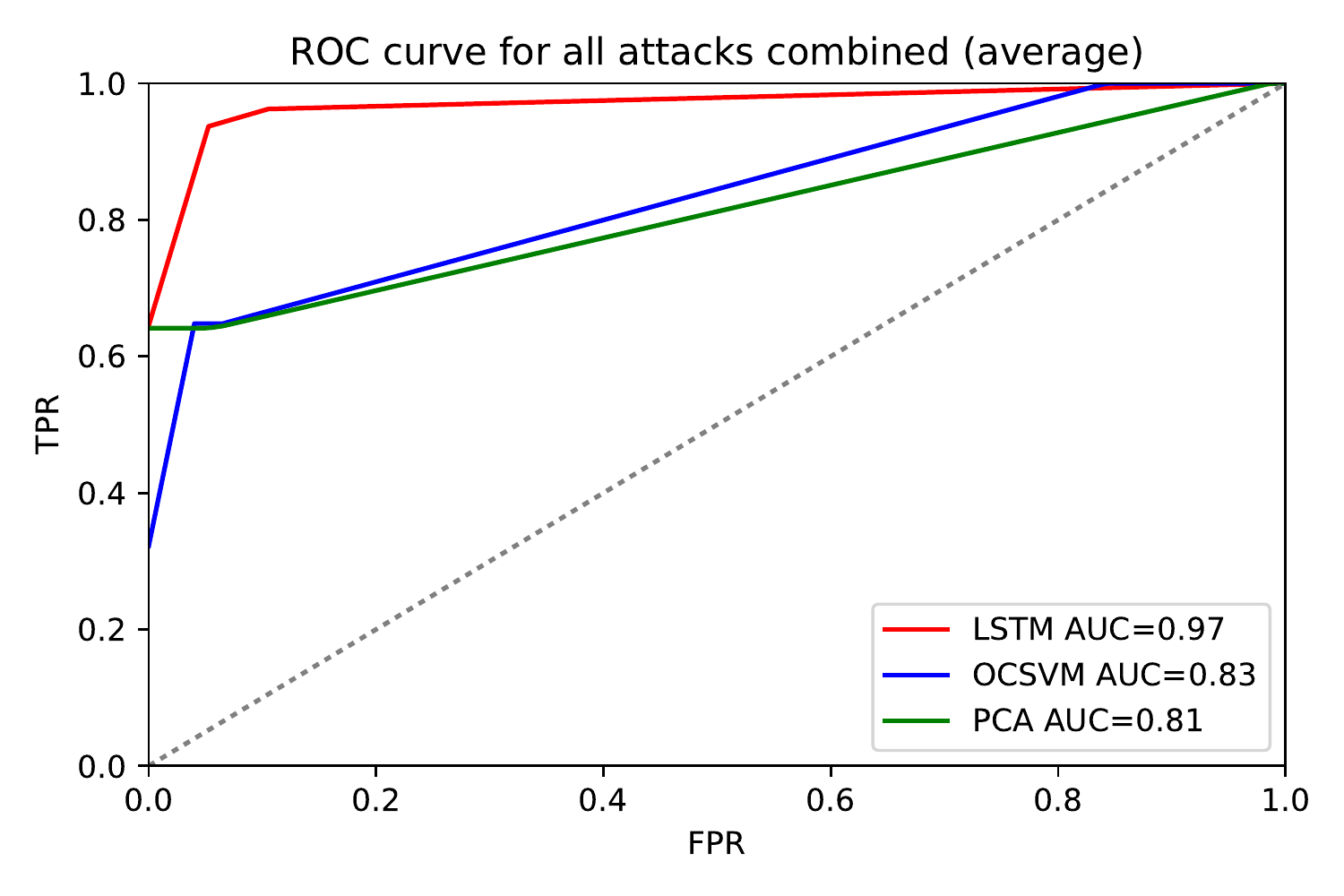}
         \caption{Struts CVE-2017-5638}
         \label{fig:struts5638_allrocs}
\end{subfigure}

\medskip
\begin{subfigure}{0.3\textwidth}
         \centering
         \includegraphics[width=\linewidth]{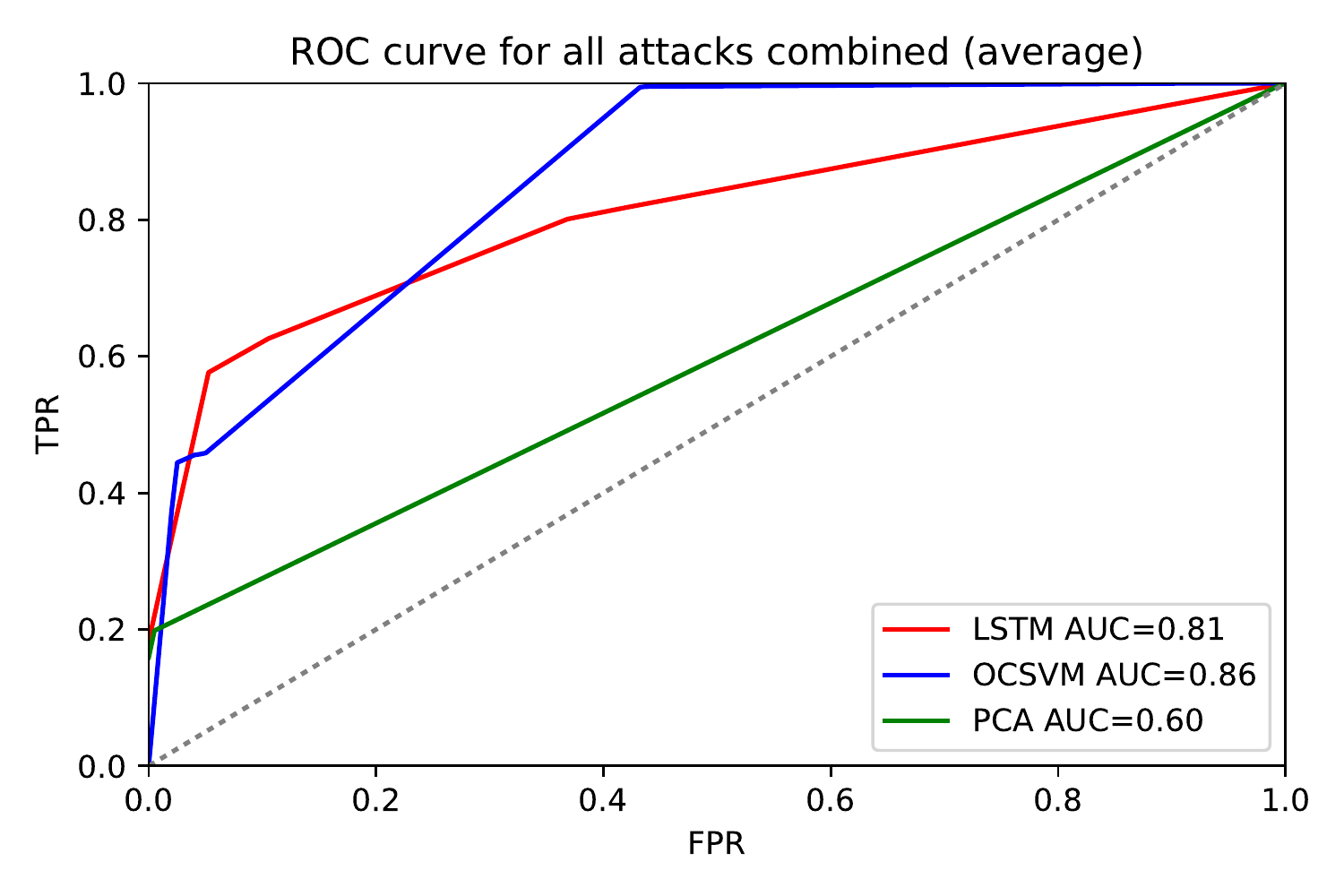}
         \caption{WP Ajax Load More Plugin}
         \label{fig:wpajax_allrocs}
\end{subfigure}
\hfill
\begin{subfigure}{0.3\textwidth}
         \centering
         \includegraphics[width=\linewidth]{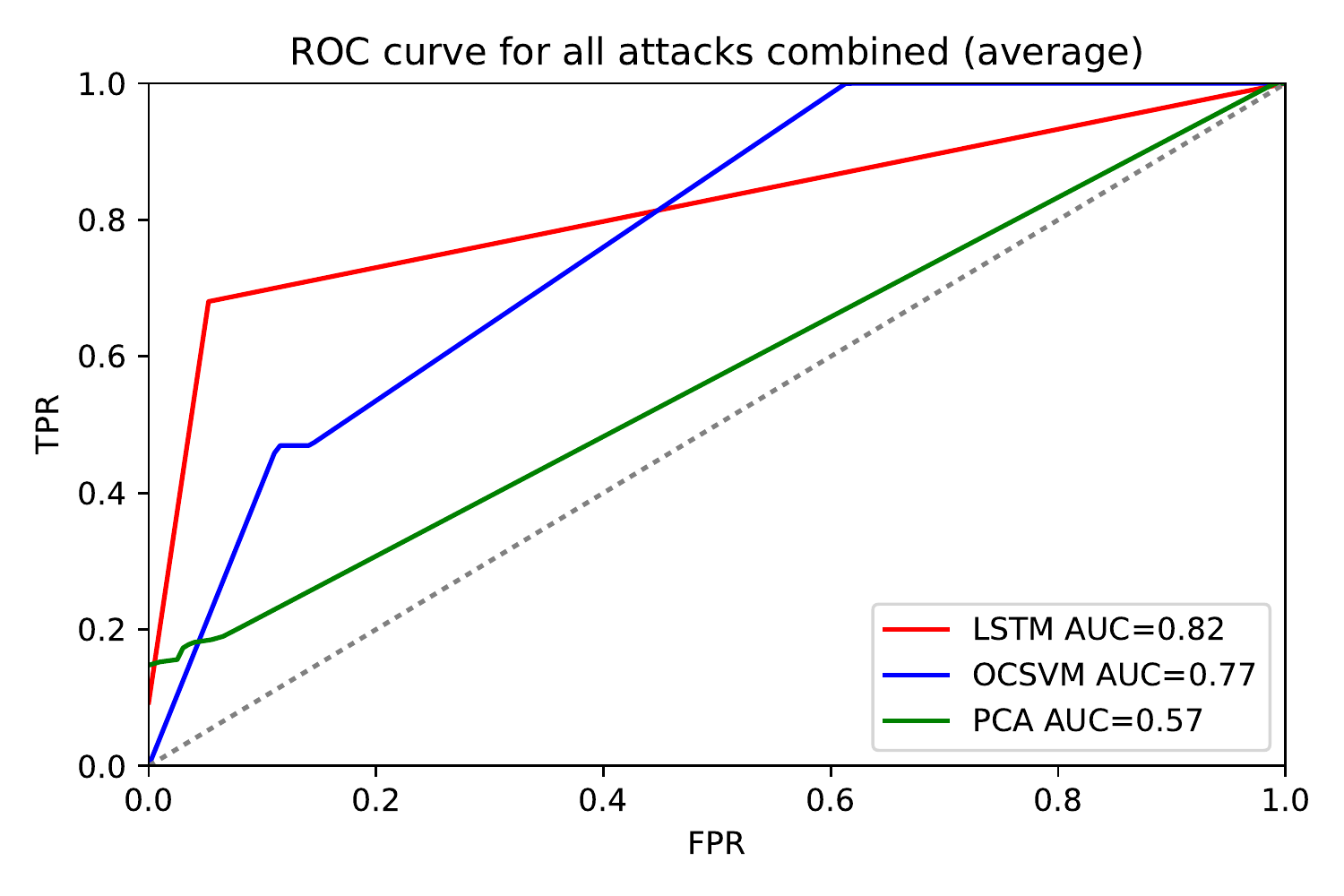}
         \caption{WP N-media Contact Form Plugin}
         \label{fig:wpnmedia_allrocs}
\end{subfigure}
\hfill
\begin{subfigure}{0.3\textwidth}
         \centering
         \includegraphics[width=\linewidth]{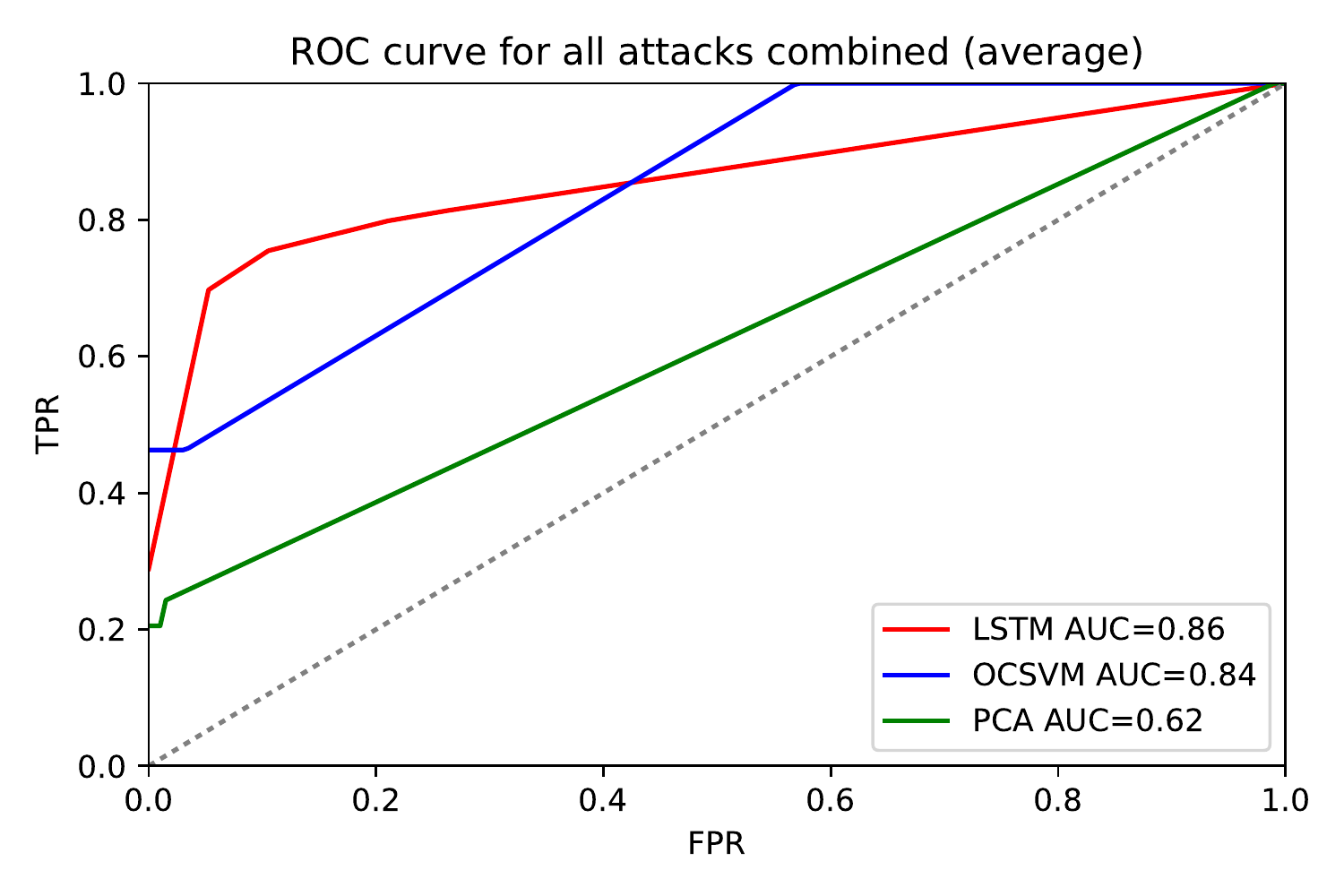}
         \caption{WP ReflexGallery Plugin}
         \label{fig:wpreflex_allrocs}
\end{subfigure}

\medskip
\begin{subfigure}{0.3\textwidth}
         \centering
         \includegraphics[width=\linewidth]{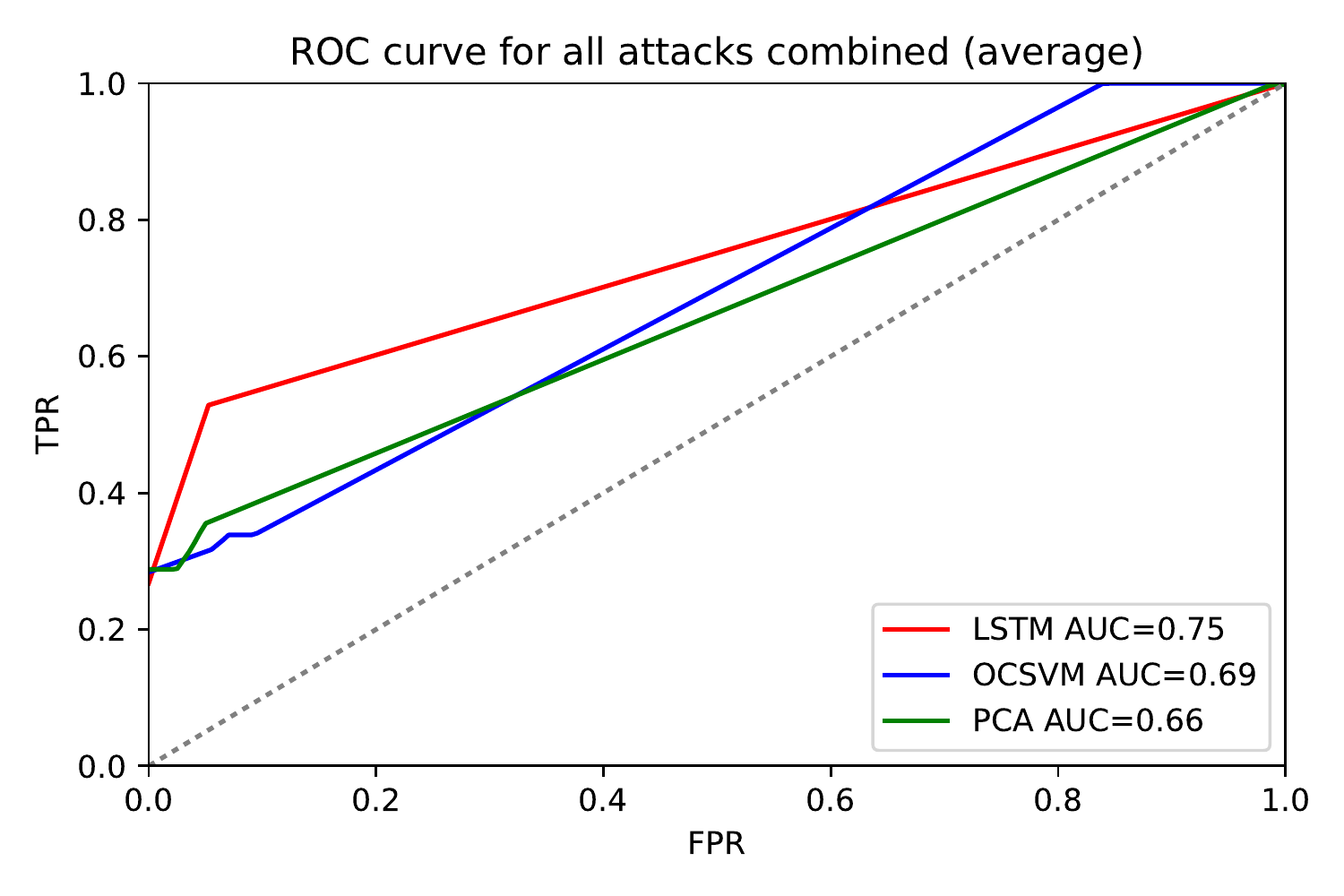}
         \caption{Struts CVE-2017-9805}
         \label{fig:struts9805_allrocs}
\end{subfigure}
\caption{ROC curves for all 15 attack scripts (averaged) for PCA, OCSVM and LSTM for all applications.}
\label{fig:allrocs}
\end{figure*}

To select a threshold for detecting anomalies, we inspected the distribution of anomaly scores for the frequency vectors in our validation set (containing only legitimate data). In Figure \ref{fig:struts5638_scoredist} we show the distribution of the distance between predicted values and actual frequency values for the Struts application with the $\mathsf{enum\_config}$ script. We show the distance for attack and legitimate system calls, when using either uniform weights, or TF-IDF weights for distance computation. As observed, when using uniform weights the two distance distribution are closer and overlap in some cases. However, with TF-IDF weights, the separation between the two distance distributions increases, making the attack data much more distinguishable from the legitimate one. Thus, we can select a threshold per application to minimize the False Positive rate during training. 



\paragraph{Comparison of LSTM with traditional models}

To compare LSTM with the two traditional anomaly detection models (PCA and OCSVM), Figure \ref{fig:strutsenumsystem} shows the ROC curves for one application (Struts) for the attack script $\mathsf{enum\_network}$. We observe that LSTM is performing significantly better, with AUC at 0.97, compared to 0.62 for PCA and 0.67 for OCSVM. In the Appendix, we show in Figure~\ref{fig:strutsrocs} the ROC curves for all attack scripts for the Struts application. Interestingly, LSTM is always outperforming the traditional models, with significant increase in AUC. For instance, for the $\mathsf{ecryptfs\_cred}$ script, LSTM achieves an AUC of 0.96, while PCA and OCSVM have AUCs of only 0.56 and 0.63, respectively.

\begin{figure*}[t]
\centering
\begin{subfigure}{.3\textwidth}
	\includegraphics[width=\linewidth]{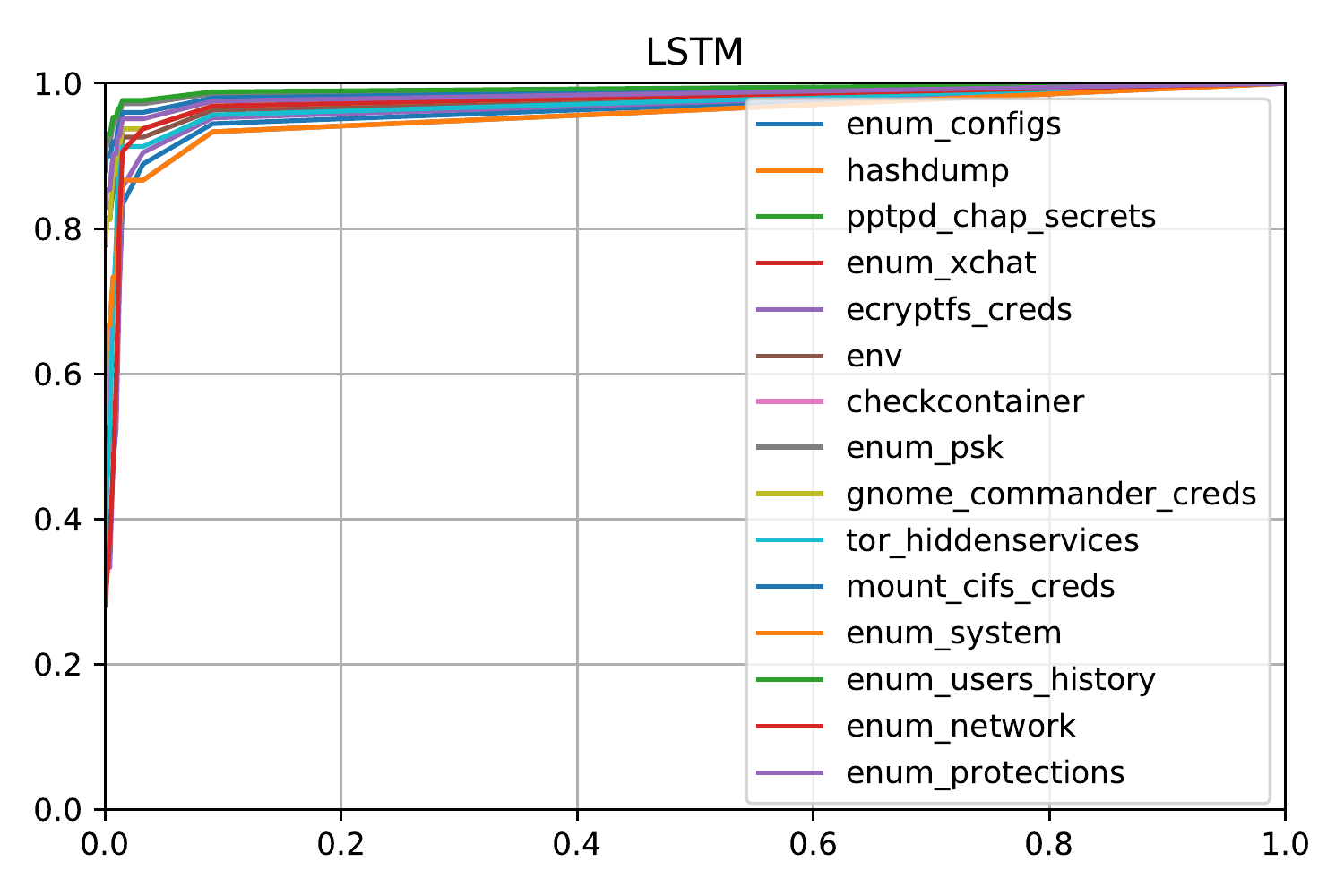}
\end{subfigure}
\begin{subfigure}{.3\textwidth}
	\includegraphics[width=\linewidth]{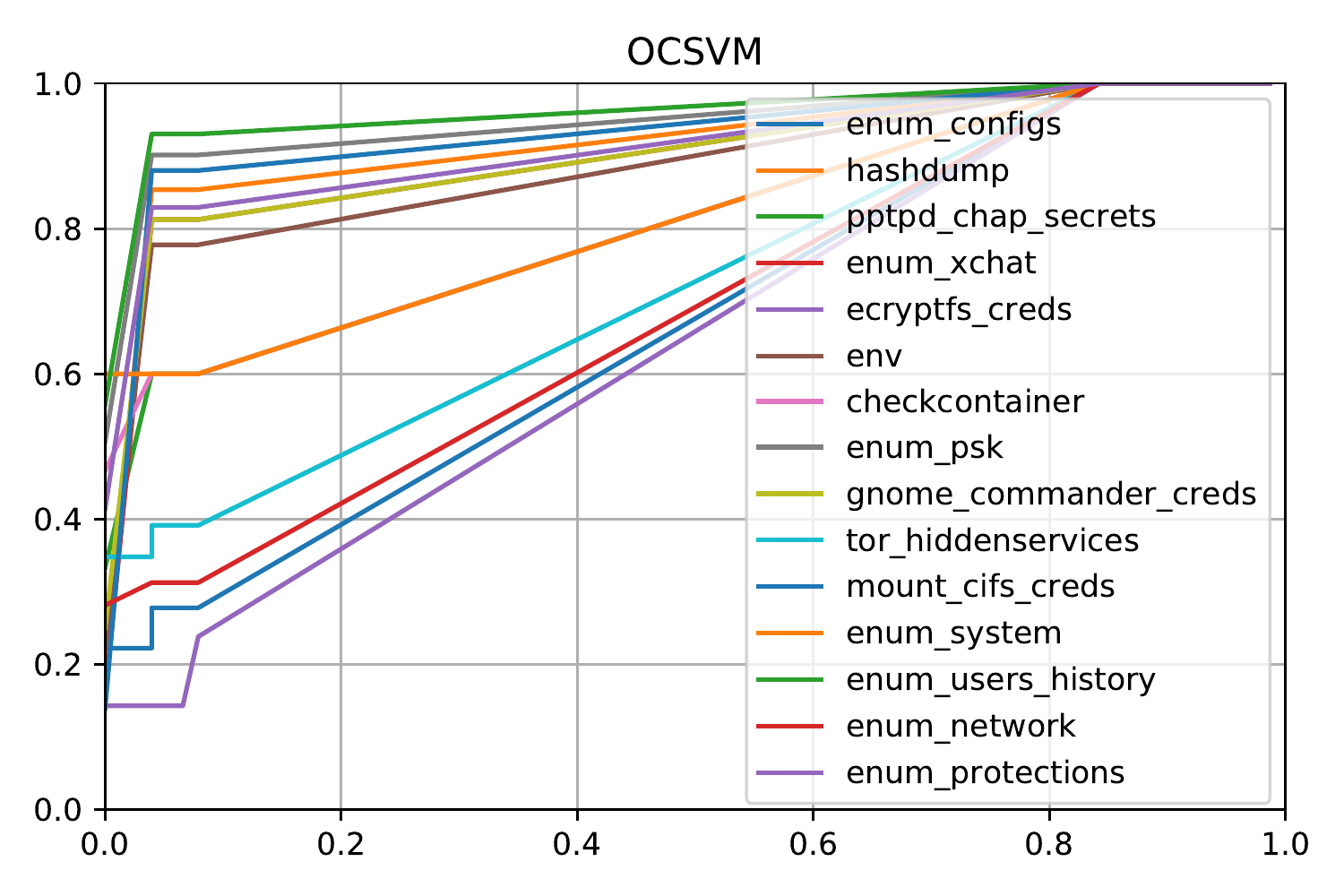}
\end{subfigure}
\begin{subfigure}{.3\textwidth}
	\includegraphics[width=\linewidth]{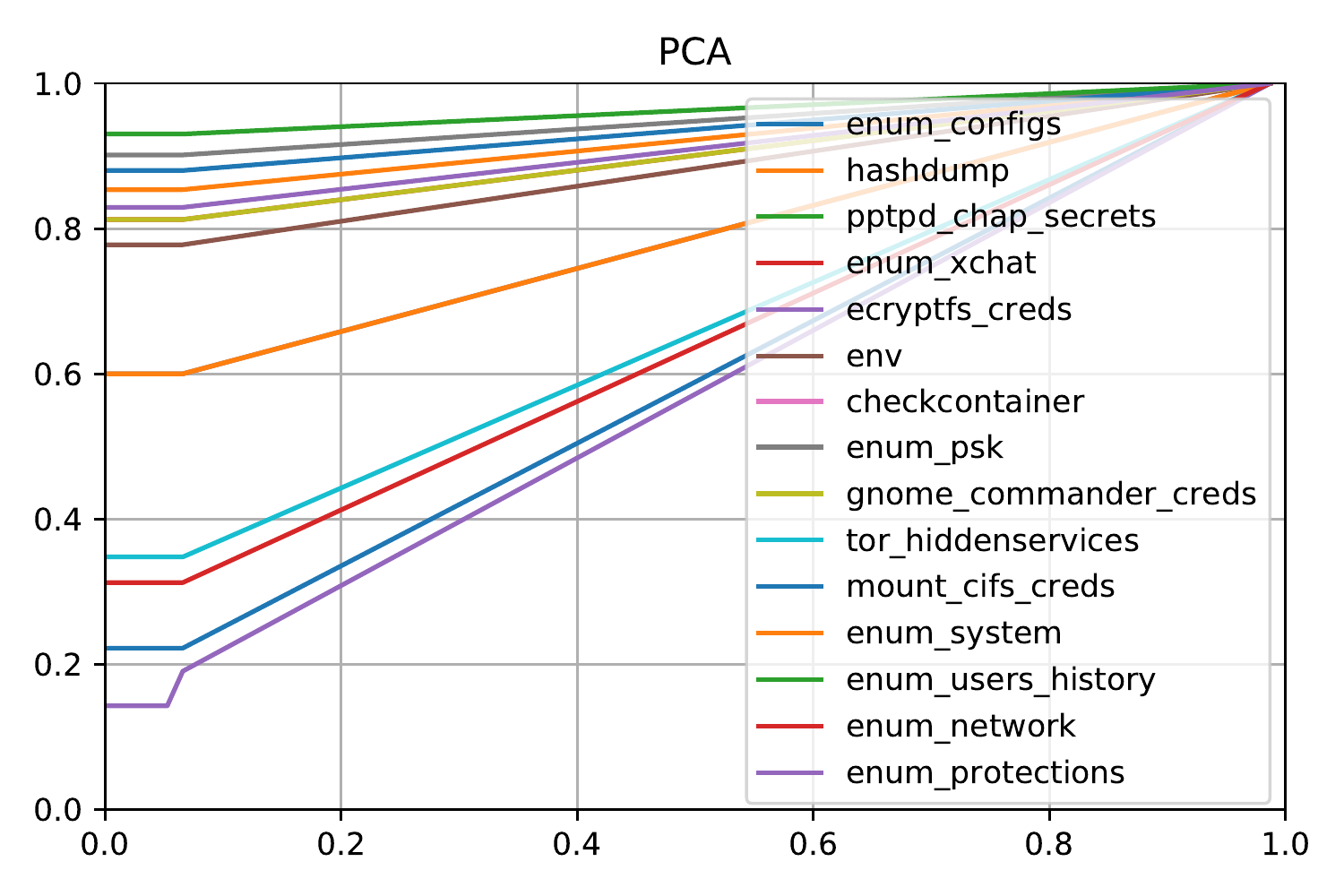}
\end{subfigure}
\caption{ROC curves for Struts with CVE-2017-5638 for all three models.}
\label{fig:struts5638allrocsseparate}
\end{figure*}

\begin{figure*}[ht]
\centering
\begin{subfigure}{.3\textwidth}
	\includegraphics[width=1.0\linewidth]{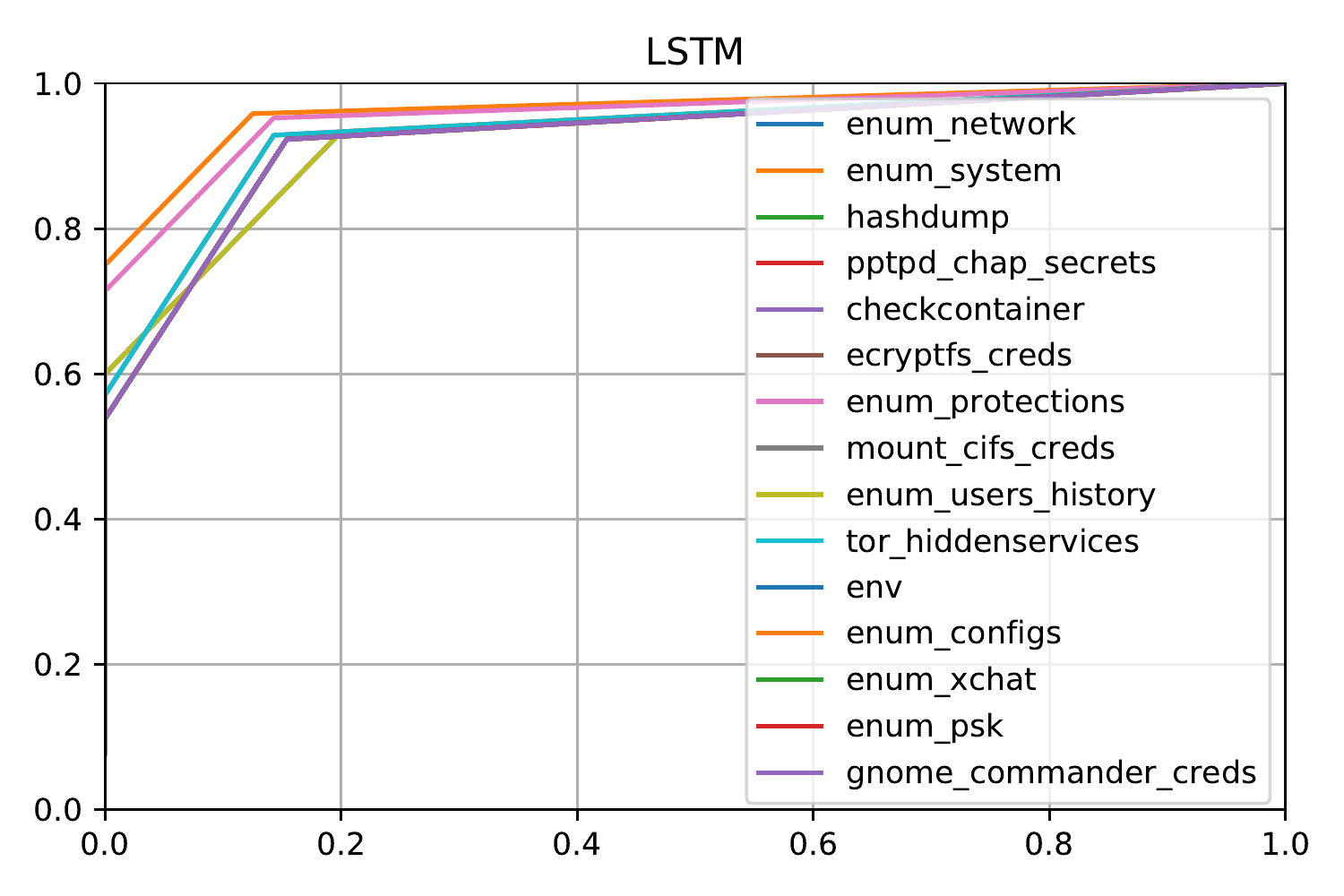}
\end{subfigure}
\begin{subfigure}{.3\textwidth}
	\includegraphics[width=1.0\linewidth]{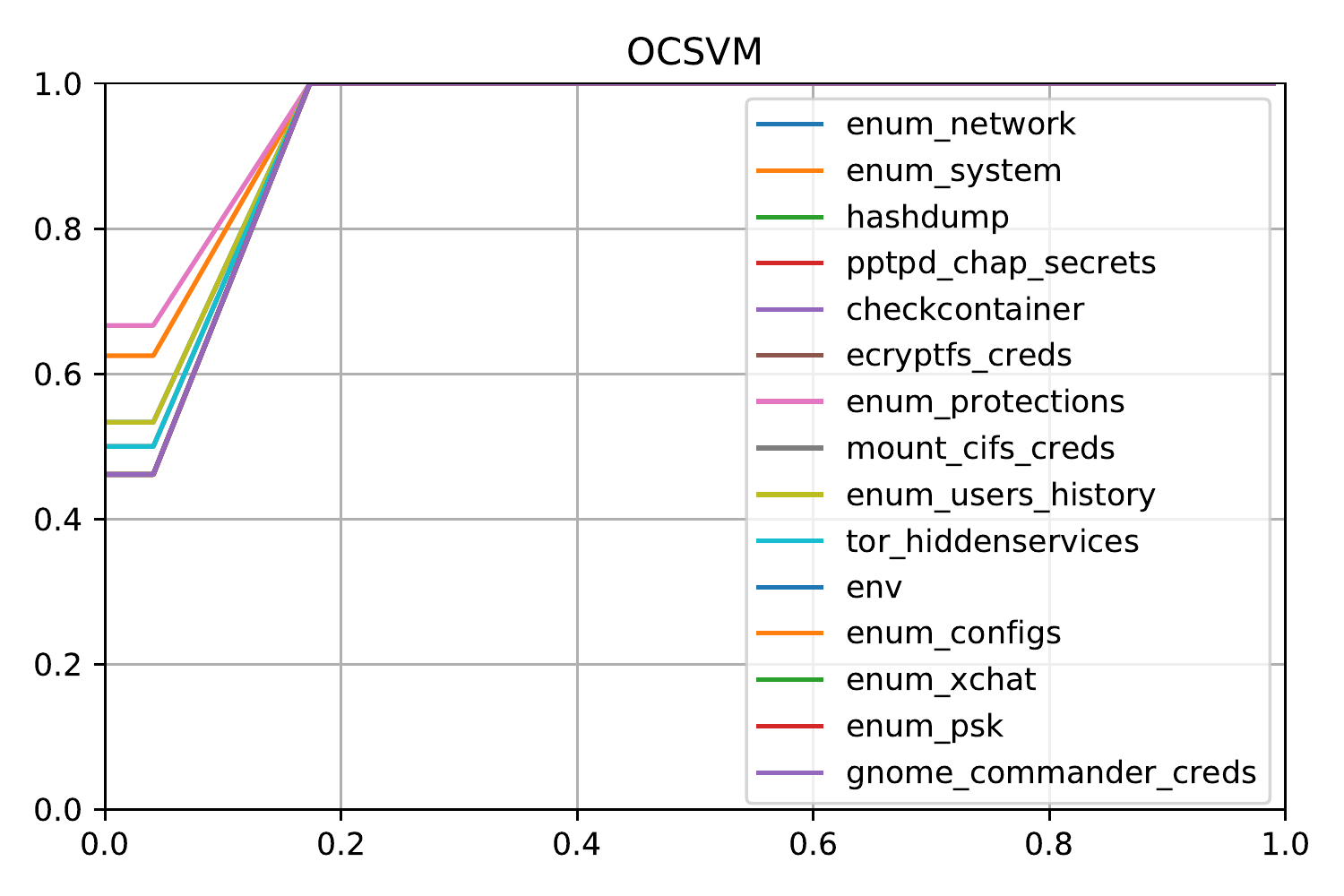}
\end{subfigure}
\begin{subfigure}{.3\textwidth}
	\includegraphics[width=1.0\linewidth]{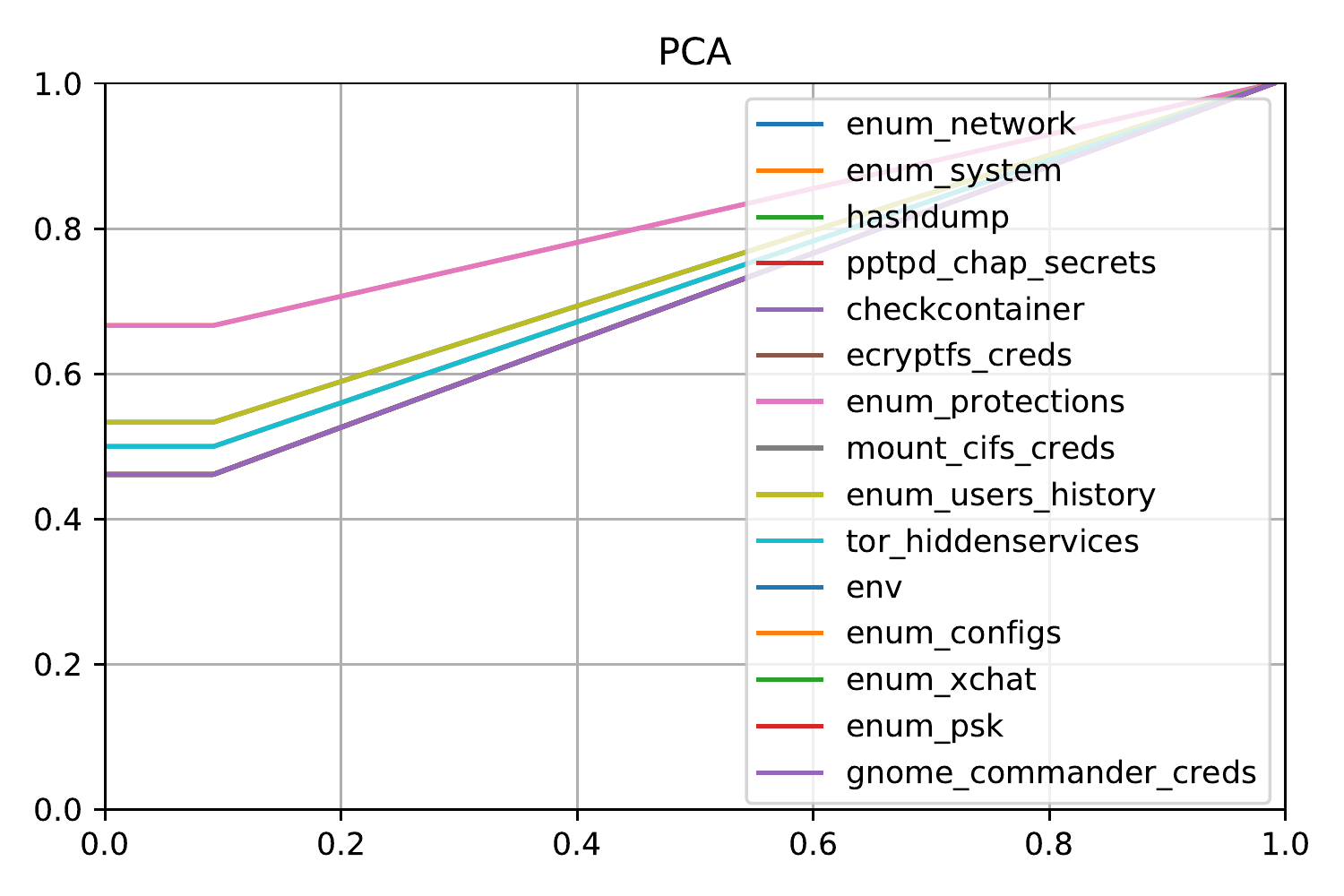}
\end{subfigure}
\caption{ROC curves for Drupal for all three models.}
\label{fig:drupalallrocsseparate}
\end{figure*}

\ignore{
\begin{figure}
\centering
\includegraphics[width=0.8\linewidth]{Figures/drupalcombined/drupal_combined_enum_system.pdf}
\caption{ROC curves for $\mathsf{enum\_system}$ for Drupal with CVE-2018-7600.}
\label{fig:drupalcombinedenumsystem}
\end{figure}
}

Figure \ref{fig:allrocs} shows the ROC curves averaged on all 15 attack scripts for the all the tested web applications. It is clear that the LSTM model performs better than the traditional models. The AUC for LSTM is between 0.75 and 0.97 and improves the traditional models' average AUC between 0.09 and 0.25.







\paragraph{LSTM for different attacks}

Figures \ref{fig:struts5638allrocsseparate} and \ref{fig:drupalallrocsseparate} show the ROC curves for each attack for the Struts and Drupal applications, respectively. The LSTM model generally performs better than PCA or OCSVM There are a few attack scripts that perform poorly for  the traditional models (e.g., $\mathsf{ecryptfs\_creds}$, $\mathsf{enum\_configs}$, and $\mathsf{enum\_network}$), but the LSTM models perform much better. We suspect that the reason is the ability of LSTM to analyze sequences of system call frequency vectors. By looking at individual feature vectors for one time window, the patterns of attack and legitimate data might be very similar, but the sequences over multiple time windows could differ significantly.  

%% file: related.tex
\section{Related work}
\label{sec:related}

Intrusion detection based on system call monitoring on end hosts has been studied in depth in the literature. Forrest et al.~\cite{forrest1996sense} create profiles of Unix process based on sequences of system calls and detect deviations under attack. Hofmeyr et al.~\cite{hofmeyr1998intrusion} expand this by using fixed-length sequences and Hamming distances between unknown sequences and sequences in the normal database for measuring dissimilarities. Warrender et al.~\cite{warrender1999detecting} introduce threshold-based sequence time delay embedding (t-STIDE), which implements an anomaly scoring technique where the score of a test sequence depends on the number of anomalous windows in the test sequence. Lane and Brodley~\cite{lane1997application,lane1997sequence} use UNIX shell command sequences to build a user profile dictionary and propose various similarity measures to detect anomalous behavior. Lee et al.~\cite{lee1998data} propose an unsupervised learning method which generates association rules from training data (using RIPPER) and use these rules to detect anomalies in test data. Mutz et al.~\cite{mutz2006anomalous} use Bayesian earning methods to model system call parameters and detect anomalies in parameter values. 

Ahmet et al.~\cite{Ahmed09} propose anomaly detection models that use both spatial and temporal features extracted from Windows API calls. Markov models have been used to model the temporal sequence of system calls and detect anomalies (e.g.,~\cite{Maggi10}). System call analysis has also been used for forensic investigation~\cite{Peisert07}. More recently, anomaly detection of system logs has been applied to enterprise networks~\cite{Beehive} and cloud deployments~\cite{DeepLog}. Beehive~\cite{Beehive} uses PCA and clustering-based methods to identify hosts with anomalous behavior in an enterprise networks, while DeepLog~\cite{DeepLog} designs an LSTM models that takes into account the sequence of system log events in HDFS and OpenStack logs to identify various anomalies.

In the area of web attacks, anomaly detection methods have been applied to create profiles of application parameters under normal conditions~\cite{Kruegel05}. Bayesian networks have been used to compose multiple models to reduce false positives~\cite{Kruegel03}. Clustering anomalies has been used for more detailed attack classification~\cite{Robertson06}. Spectogram~\cite{Spectogram} designs a mixture of Markov chains based on n-gram features to model the normal behavior of HTTP request parameters. Swaddler~\cite{Swaddler} models the worflow of an application and uses anomaly detection to identify when an application reaches an inconsistent state. Scarcity of training data for client web application has been mitigated by Robertson et al.~\cite{Robertson10} by using global similarity of web requests. Several papers~\cite{Shar15,Medeiros15} use a hybrid approach based on program analysis and machine learning for web application vulnerability detection. Mutz et al.~\cite{Mutz07} use a combination of dynamic analysis and learning to determine the context of system call invocation and identify anomalous sequences of system calls. 

Other defenses against Web application vulnerabilities are based on either: (1) static analysis (e.g.,~\cite{Pixy}); (2) dynamic analysis (e.g.,~\cite{XSSDynamic}); (3) combination of static and dynamic analysis (e.g.,~\cite{Saner}); (3) input validation (e.g.,~\cite{Scholte12}); and (4) fuzz testing (e.g.,~\cite{Allister08}), but they have well-known limitations. For instance, most static analysis tools have large false positives, and input validation methods are specific to certain web attacks such as SQL injection. We believe that machine learning models can complement existing defenses in web applications deployed in either public or private clouds.




%% file: conclusions.tex
\section{Conclusions}
\label{sec:conclusions}

We proposed an anomaly detection framework for detecting exploits in web application. We set up a testbed environment in which we deployed four popular web application, recreated seven exploits using Metasploit modules, and collected system call data using the Sysdig monitoring agent. We compare two traditional anomaly detection models (PCA and OCSVM) with an LSTM-model trained on sequences of system call frequency vectors. We demonstrate that LSTM outperforms the traditional models. Our framework has the advantage of not requiring attack data for training, and being applicable to a range of web application exploits. In future work, interesting research questions apply on extending this framework to other scenarios and running the algorithms on data collected from real cloud environments.

\section*{Acknowledgements}

We would like to thank Vinny Parla, Andrew Zawadowskiy, and Donovan O'Hara  from Cisco for suggesting the area of research and providing feedback and guidance during the design and evaluation of \sys. This project was funded by a research gift from Cisco, as well as the NSF grant CNS-1717634.

%% file: appendix_short.tex
\appendix

\section{ROC curves for attacks} 
\begin{figure*}[h]
\centering
\begin{subfigure}{0.3\textwidth}
	\centering
	\includegraphics[height=0.15\textheight,keepaspectratio]{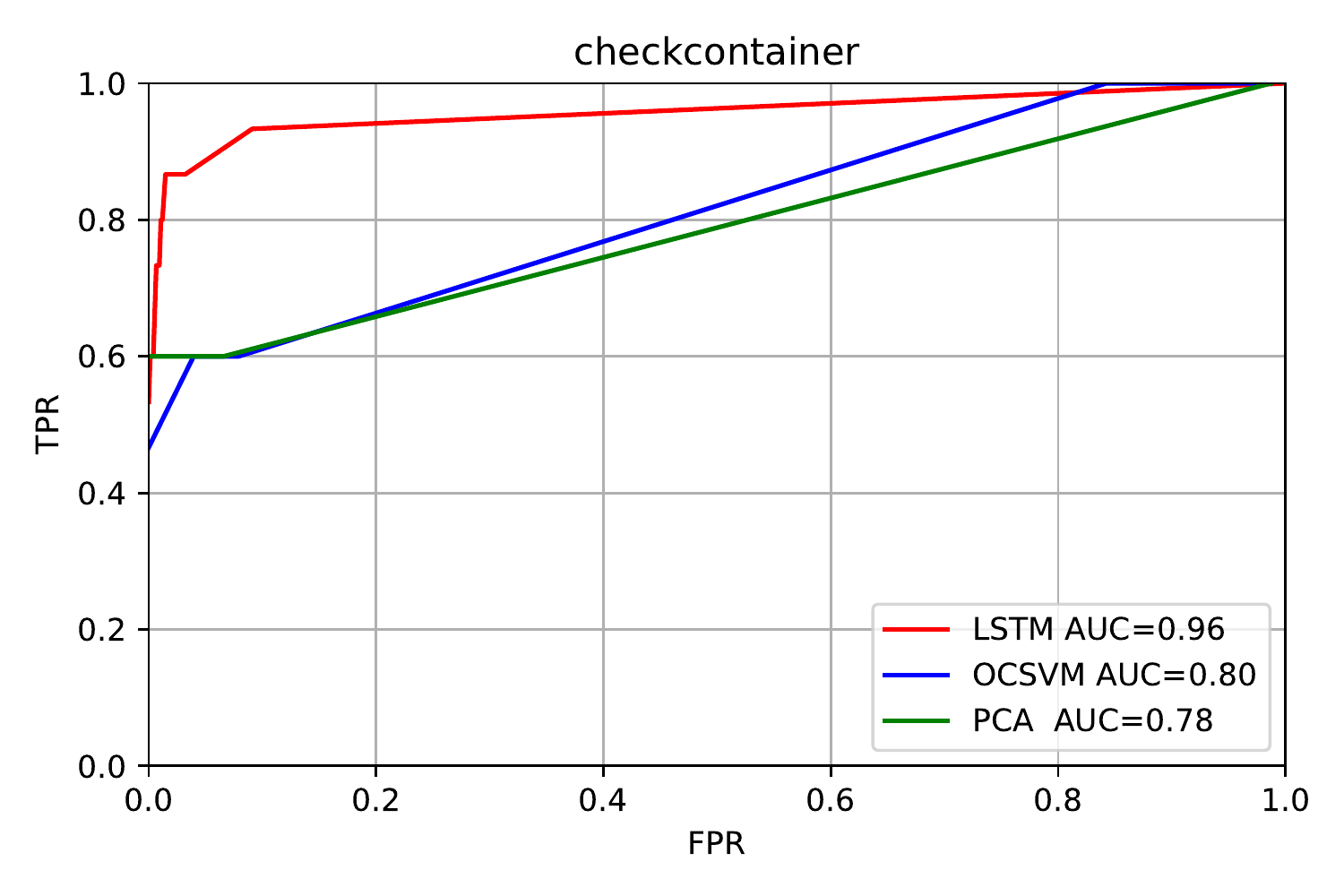}
\end{subfigure}
\hfill
\begin{subfigure}{0.3\textwidth}
         \centering
         \includegraphics[height=0.15\textheight,keepaspectratio]{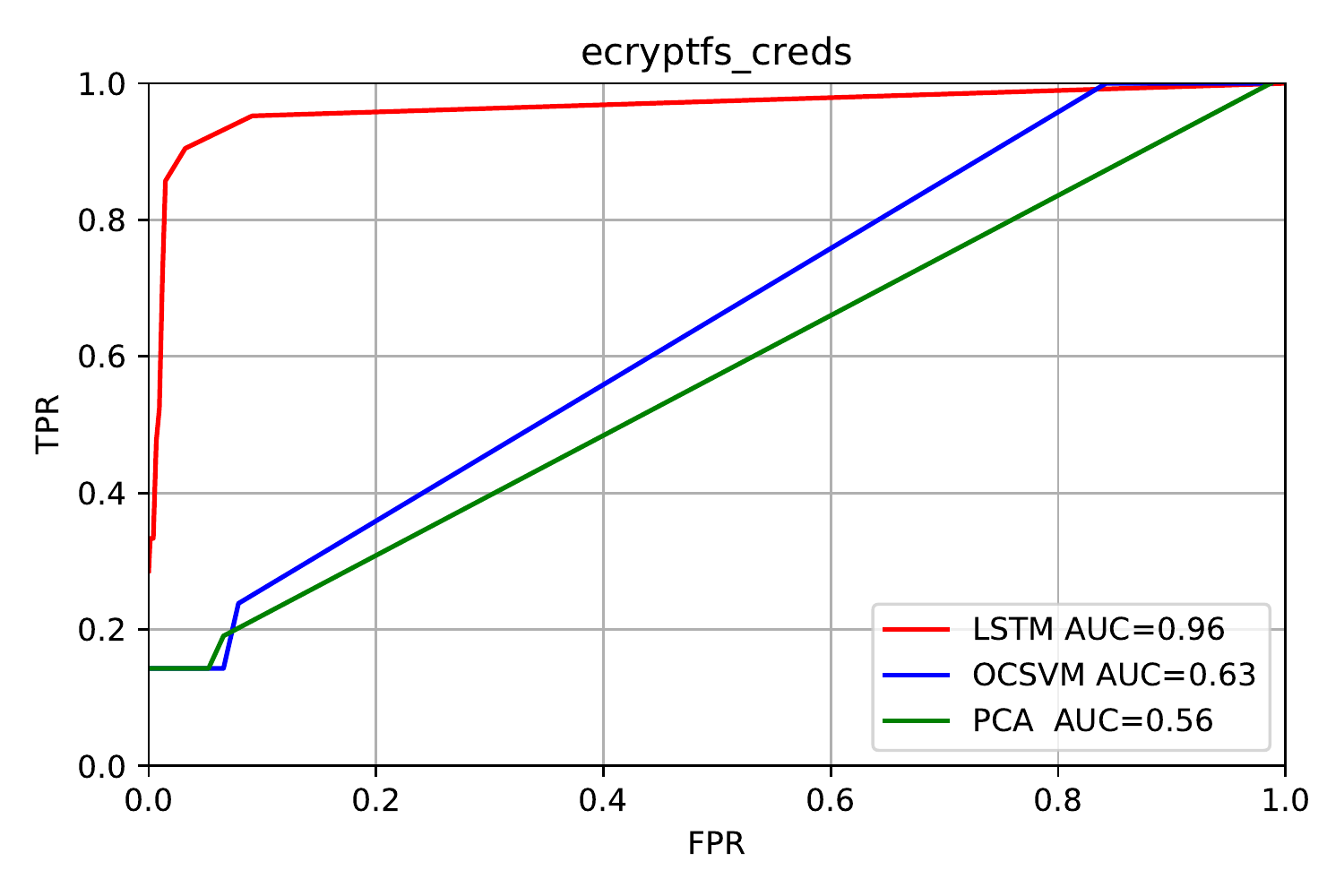}
\end{subfigure}
\hfill
\begin{subfigure}{0.3\textwidth}
        \centering
         \includegraphics[height=0.15\textheight,keepaspectratio]{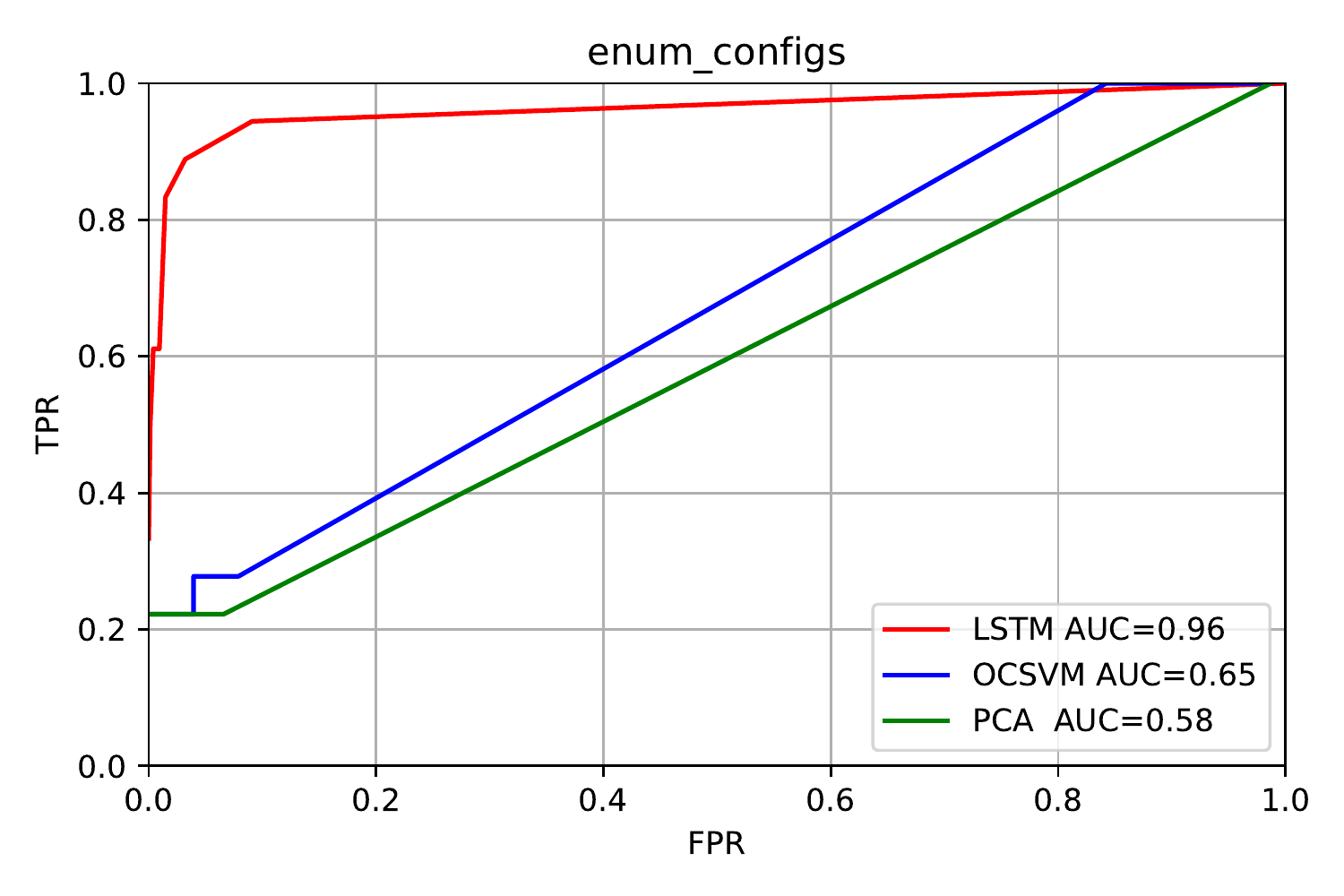}
\end{subfigure}

\smallskip
\begin{subfigure}{0.3\textwidth}
	\centering
	\includegraphics[height=0.15\textheight,keepaspectratio]{struts5638_combined_enum_network.pdf}
\end{subfigure}
\hfill
\begin{subfigure}{0.3\textwidth}
         \centering
         \includegraphics[height=0.15\textheight,keepaspectratio]{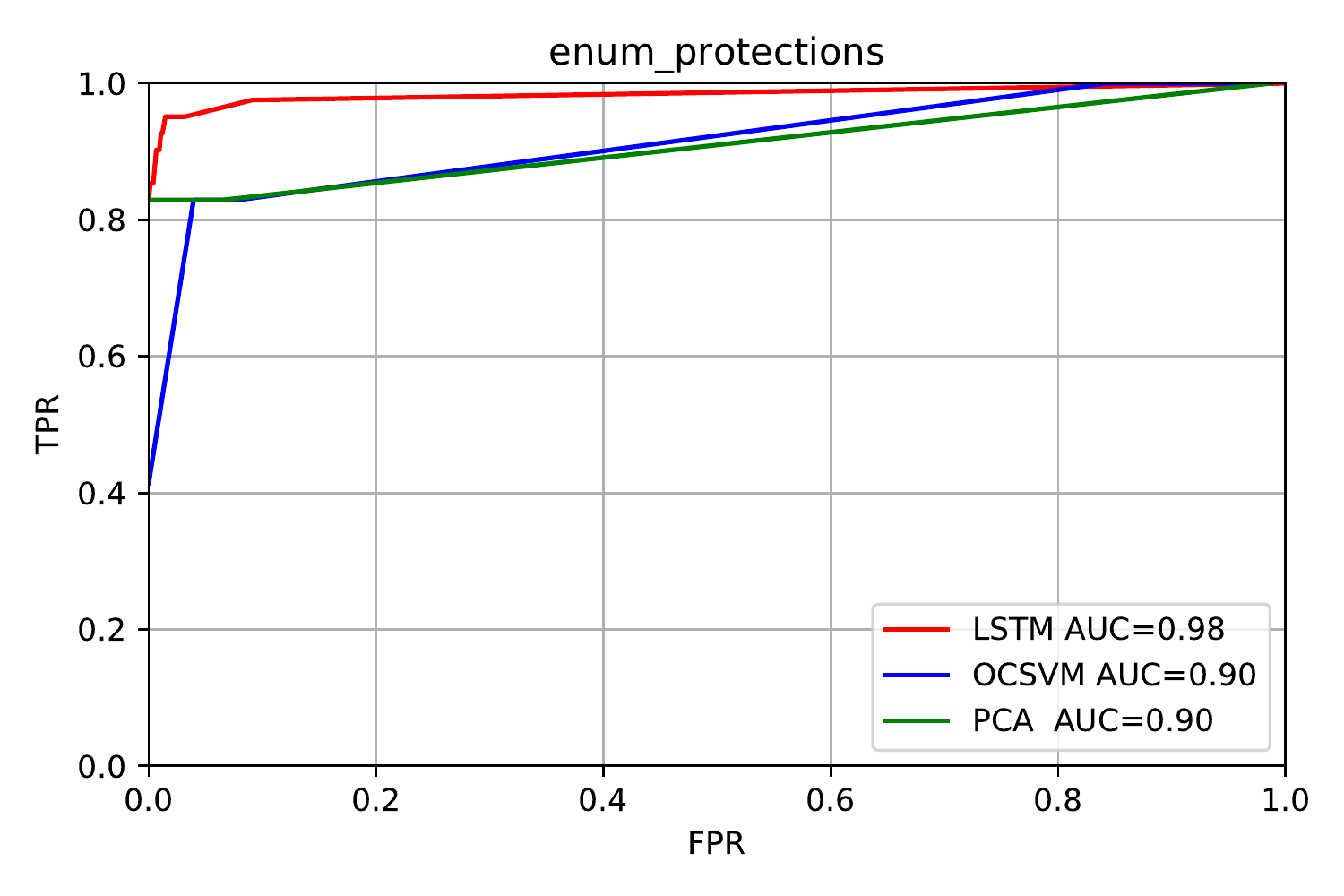}
\end{subfigure}
\hfill
\begin{subfigure}{0.3\textwidth}
        \centering
         \includegraphics[height=0.15\textheight,keepaspectratio]{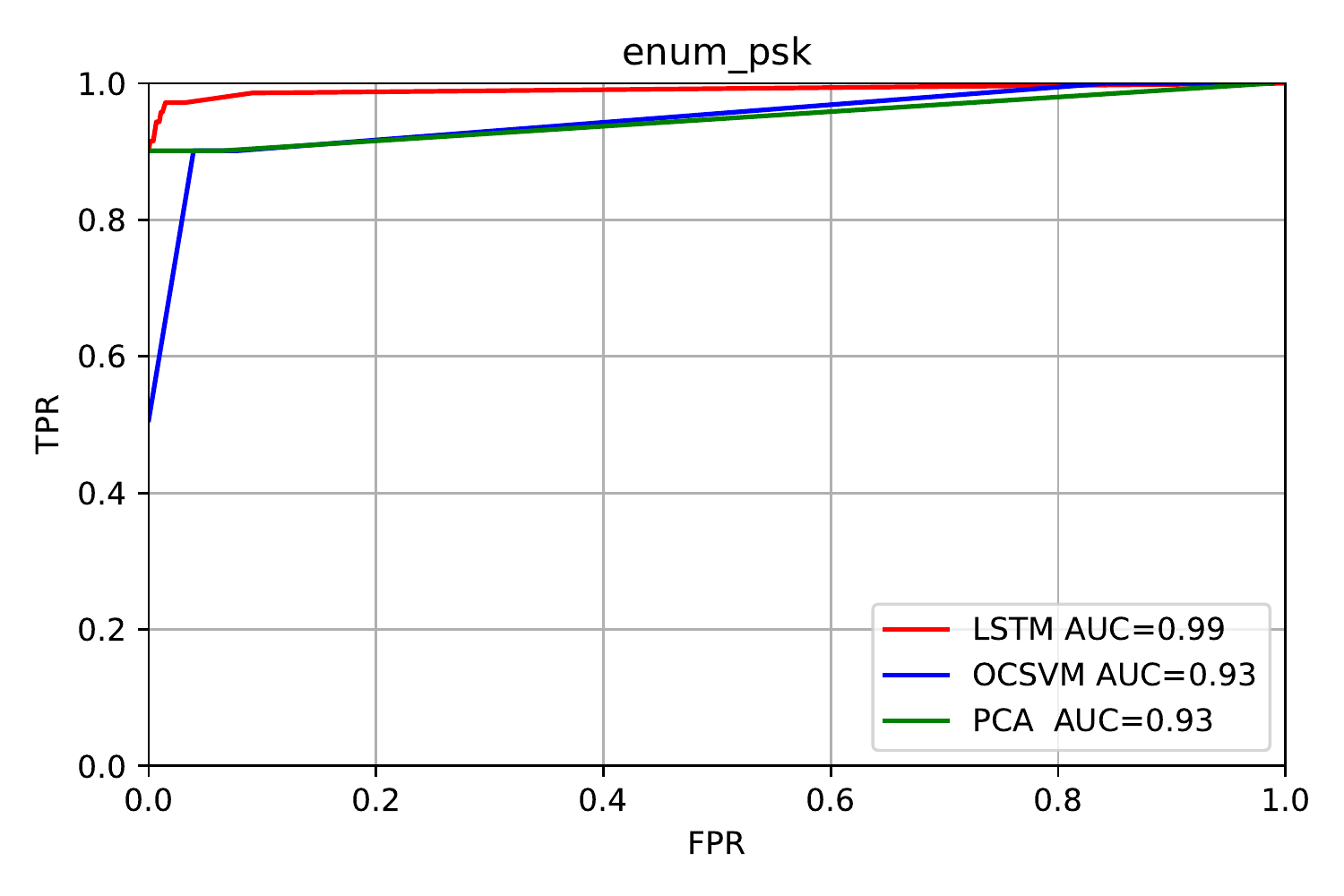}
\end{subfigure}

\smallskip
\begin{subfigure}{0.3\textwidth}
	\centering
	\includegraphics[height=0.15\textheight,keepaspectratio]{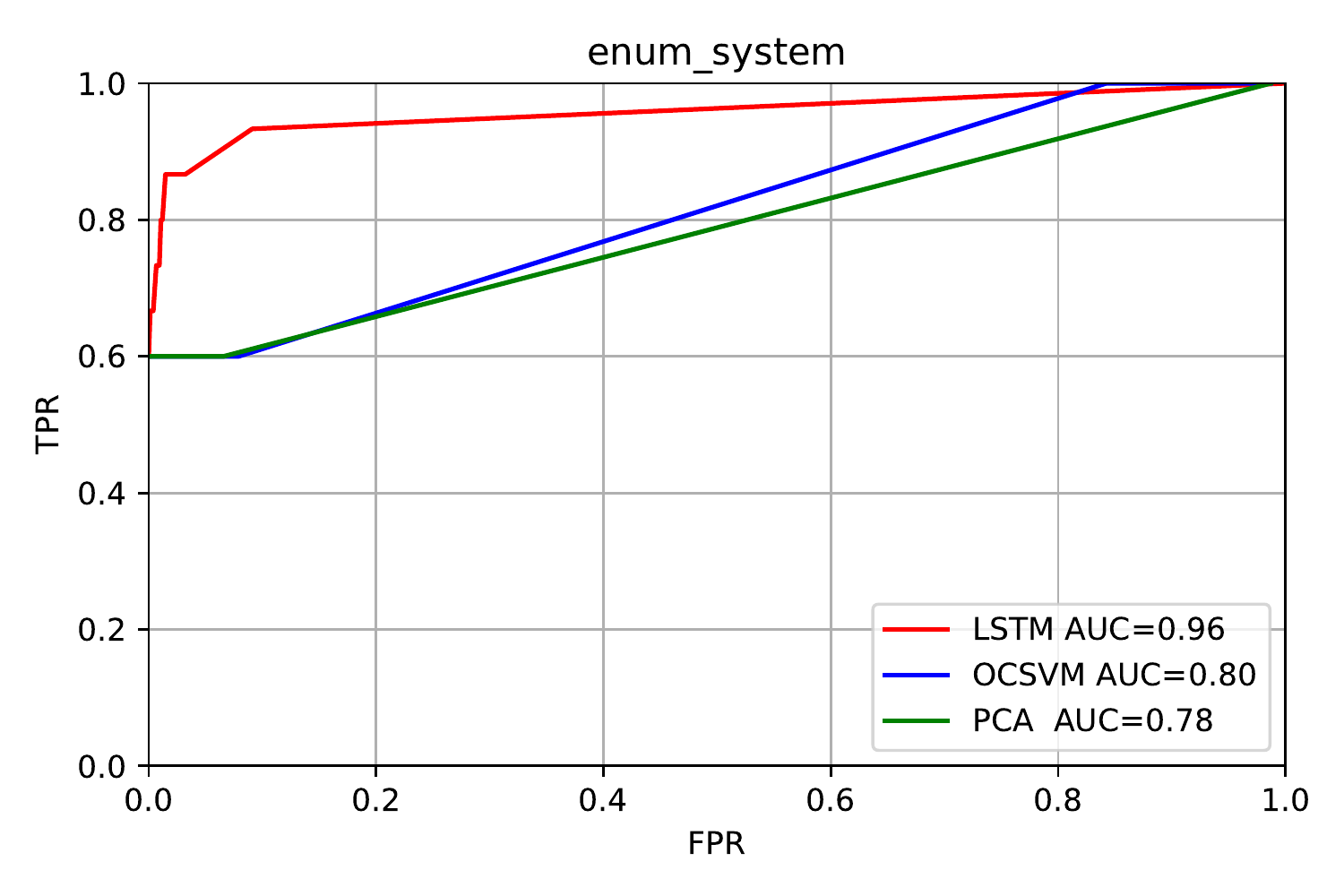}
\end{subfigure}
\hfill
\begin{subfigure}{0.3\textwidth}
         \centering
         \includegraphics[height=0.15\textheight,keepaspectratio]{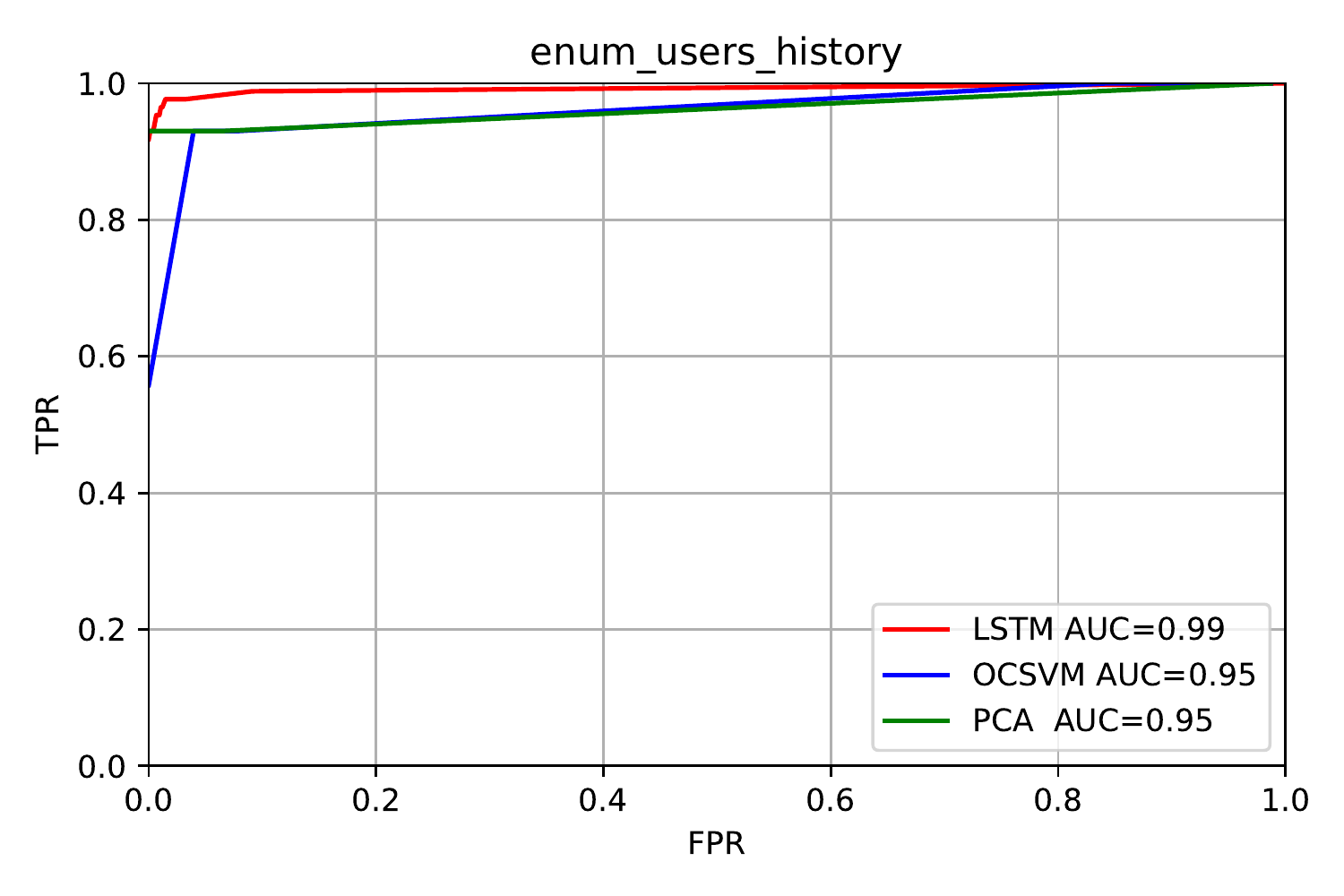}
\end{subfigure}
\hfill
\begin{subfigure}{0.3\textwidth}
        \centering
         \includegraphics[height=0.15\textheight,keepaspectratio]{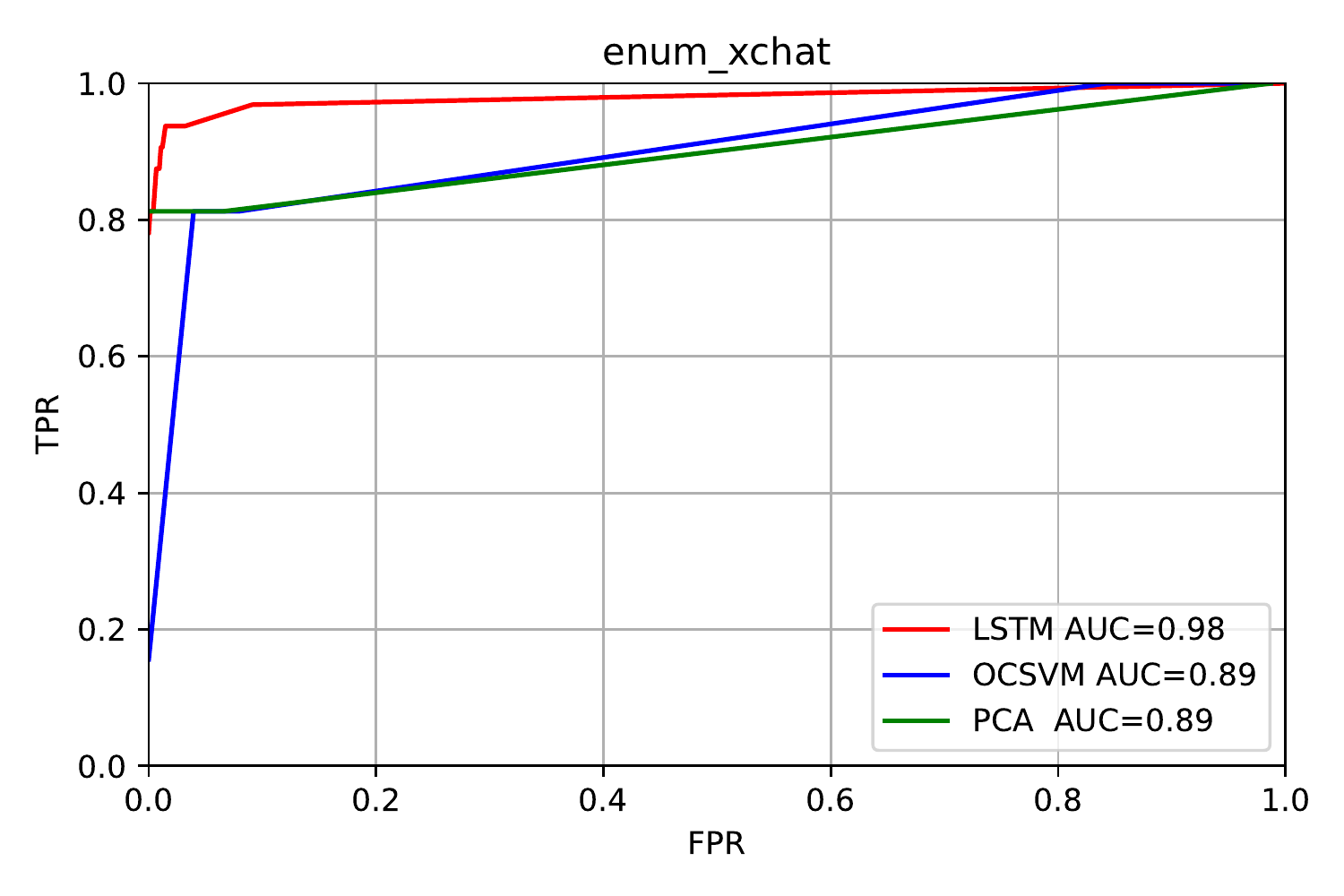}
\end{subfigure}

\smallskip
\begin{subfigure}{0.3\textwidth}
	\centering
	\includegraphics[height=0.15\textheight,keepaspectratio]{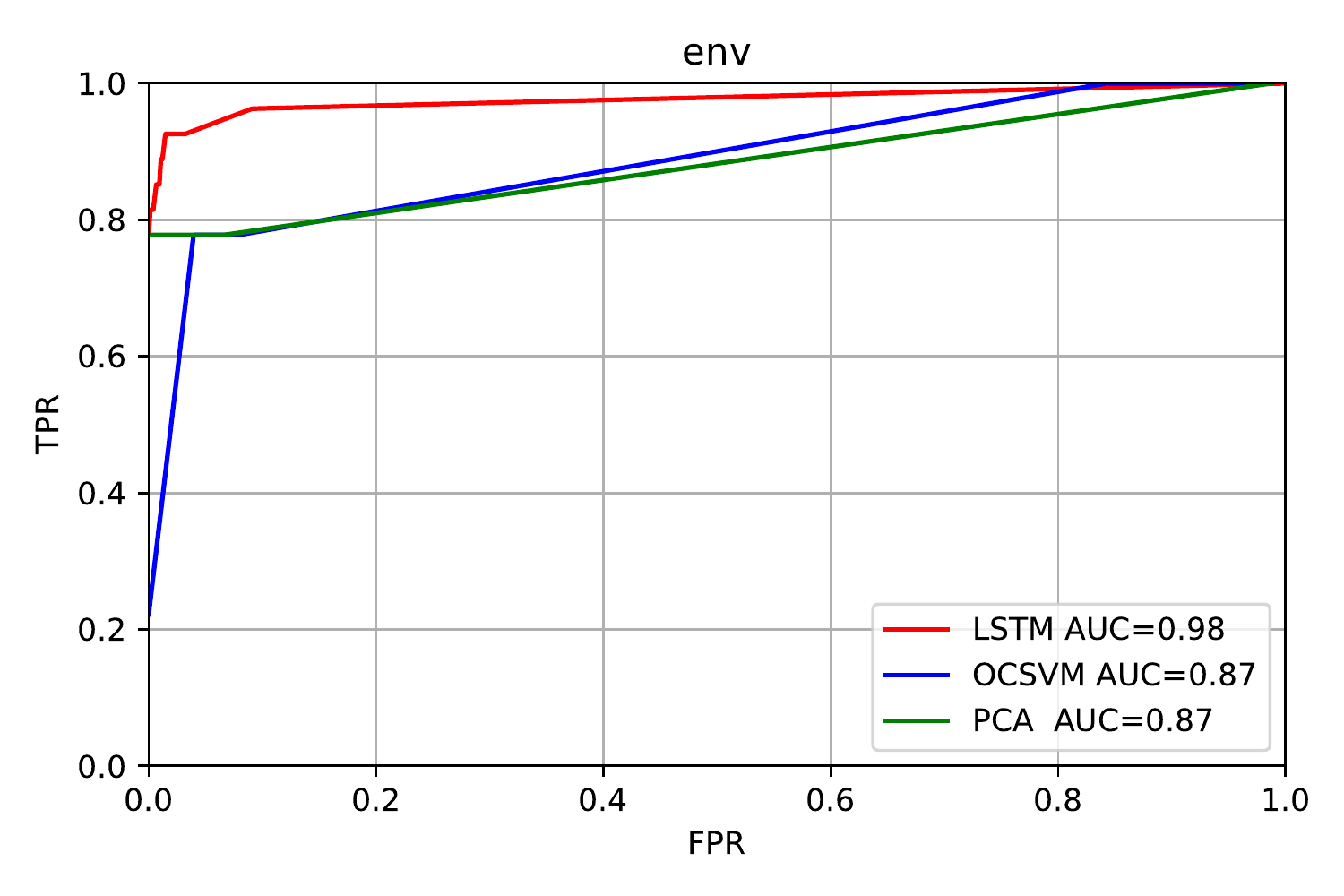}
\end{subfigure}
\hfill
\begin{subfigure}{0.3\textwidth}
         \centering
         \includegraphics[height=0.15\textheight,keepaspectratio]{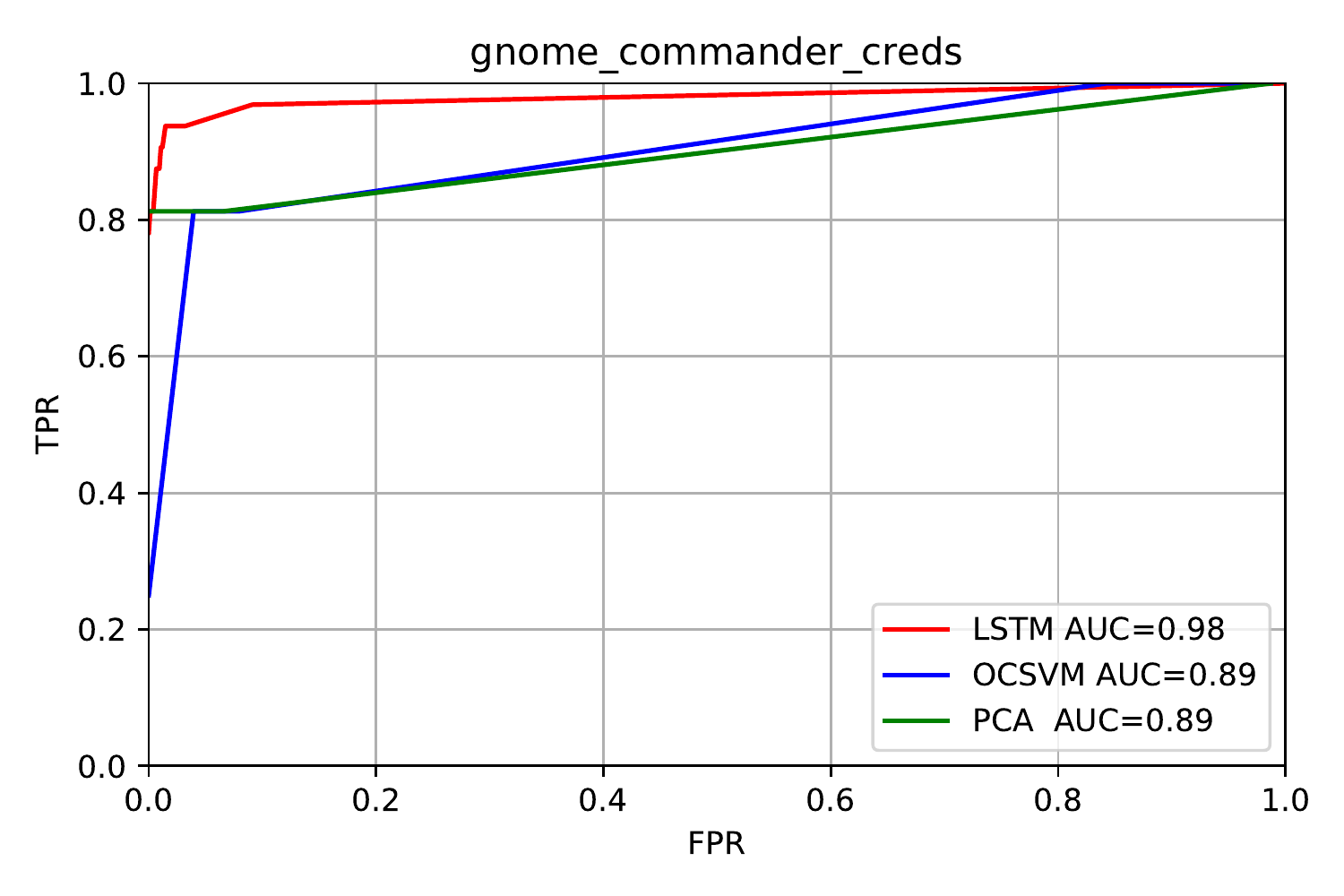}
\end{subfigure}
\hfill
\begin{subfigure}{0.3\textwidth}
        \centering
         \includegraphics[height=0.15\textheight,keepaspectratio]{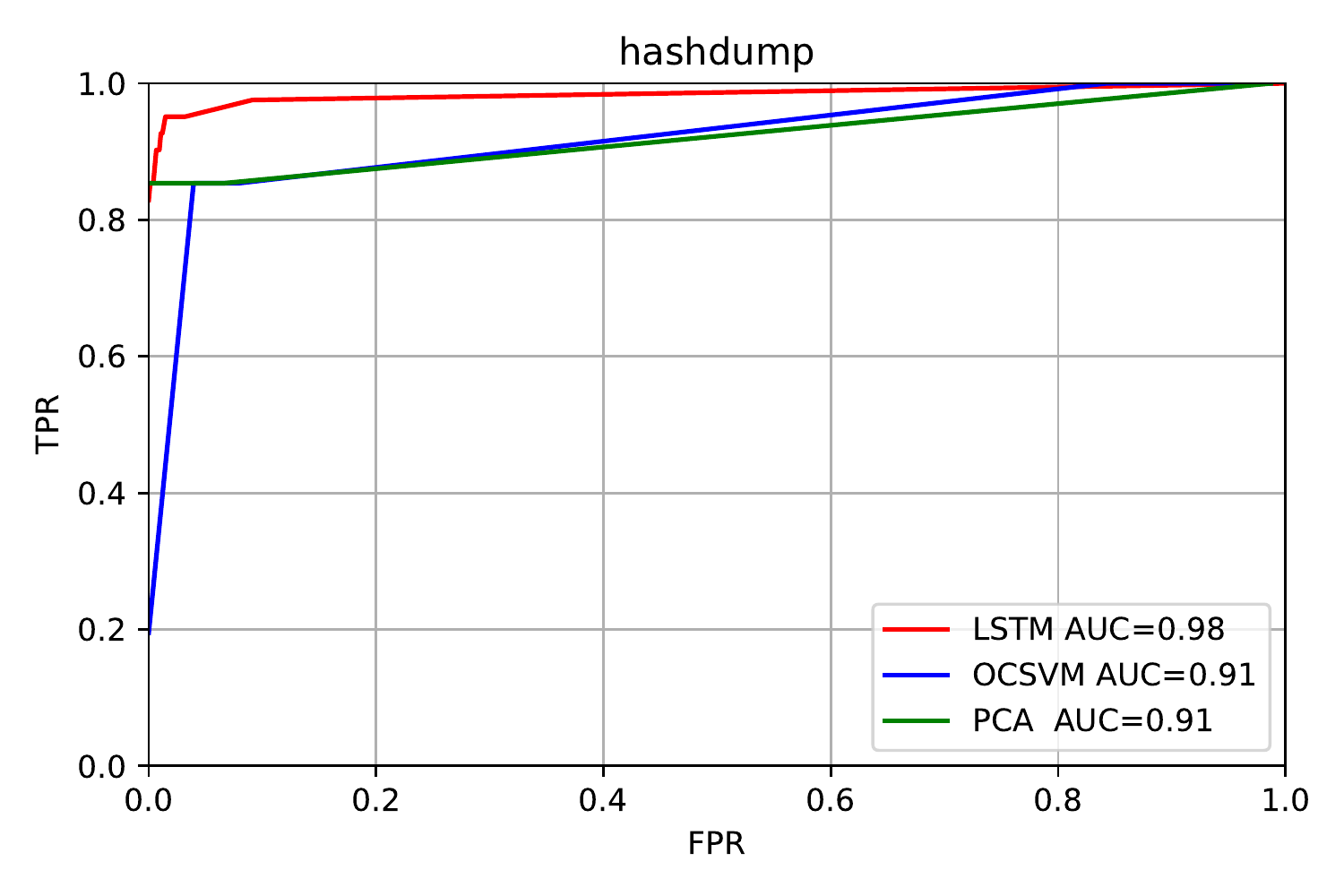}
\end{subfigure}

\smallskip
\begin{subfigure}{0.3\textwidth}
	\centering
	\includegraphics[height=0.15\textheight,keepaspectratio]{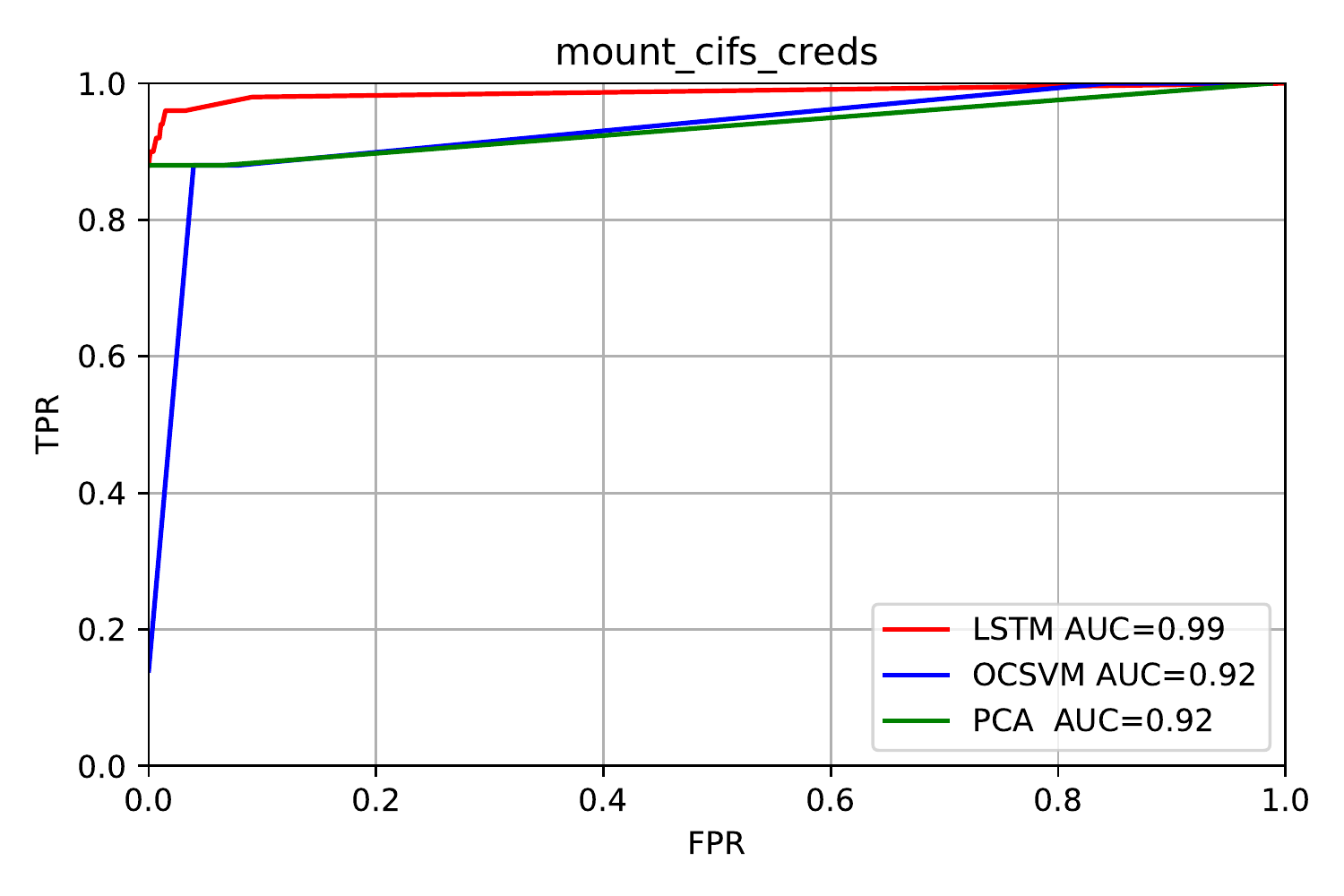}
\end{subfigure}
\hfill
\begin{subfigure}{0.3\textwidth}
         \centering
         \includegraphics[height=0.15\textheight,keepaspectratio]{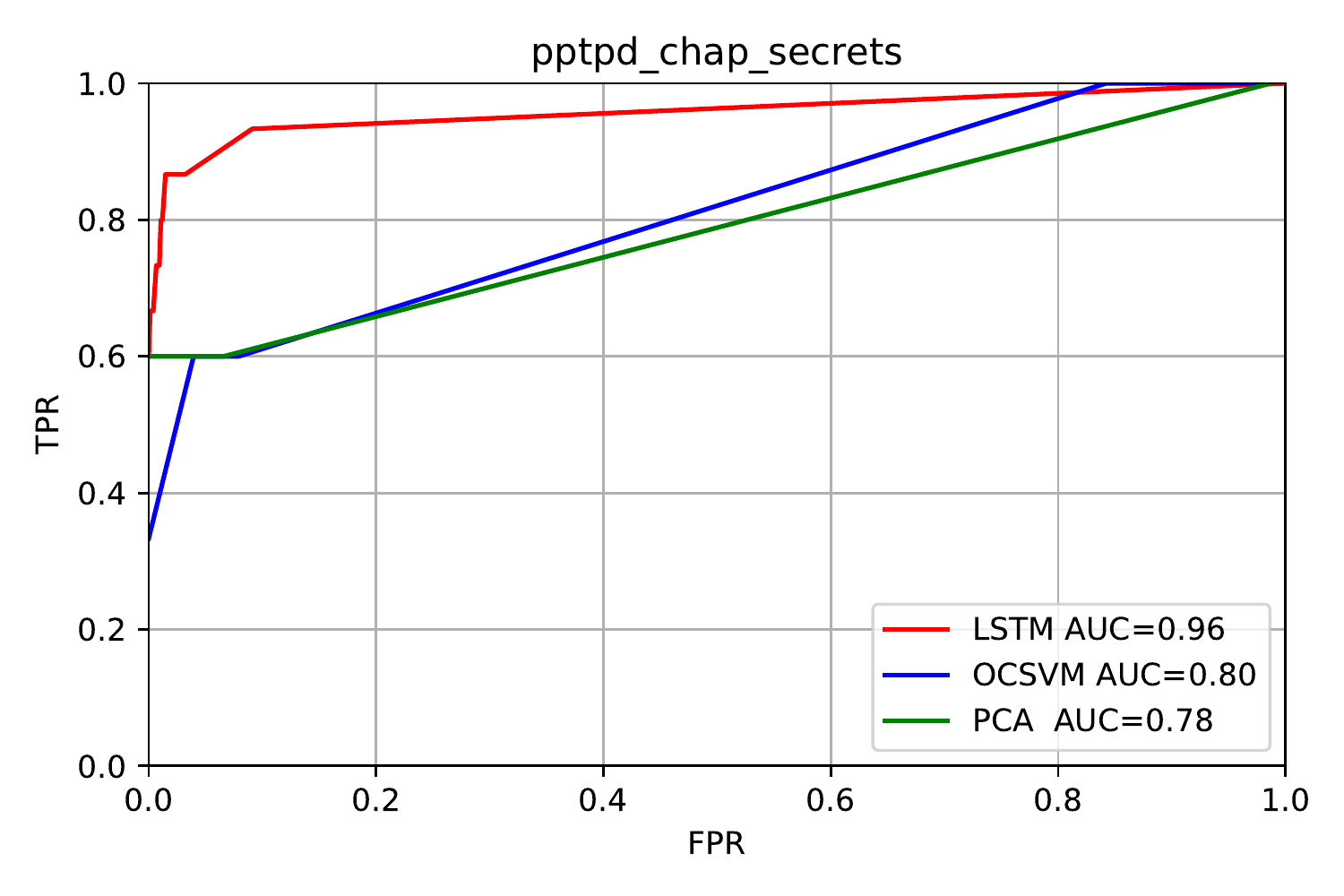}
\end{subfigure}
\hfill
\begin{subfigure}{0.3\textwidth}
        \centering
         \includegraphics[height=0.15\textheight,keepaspectratio]{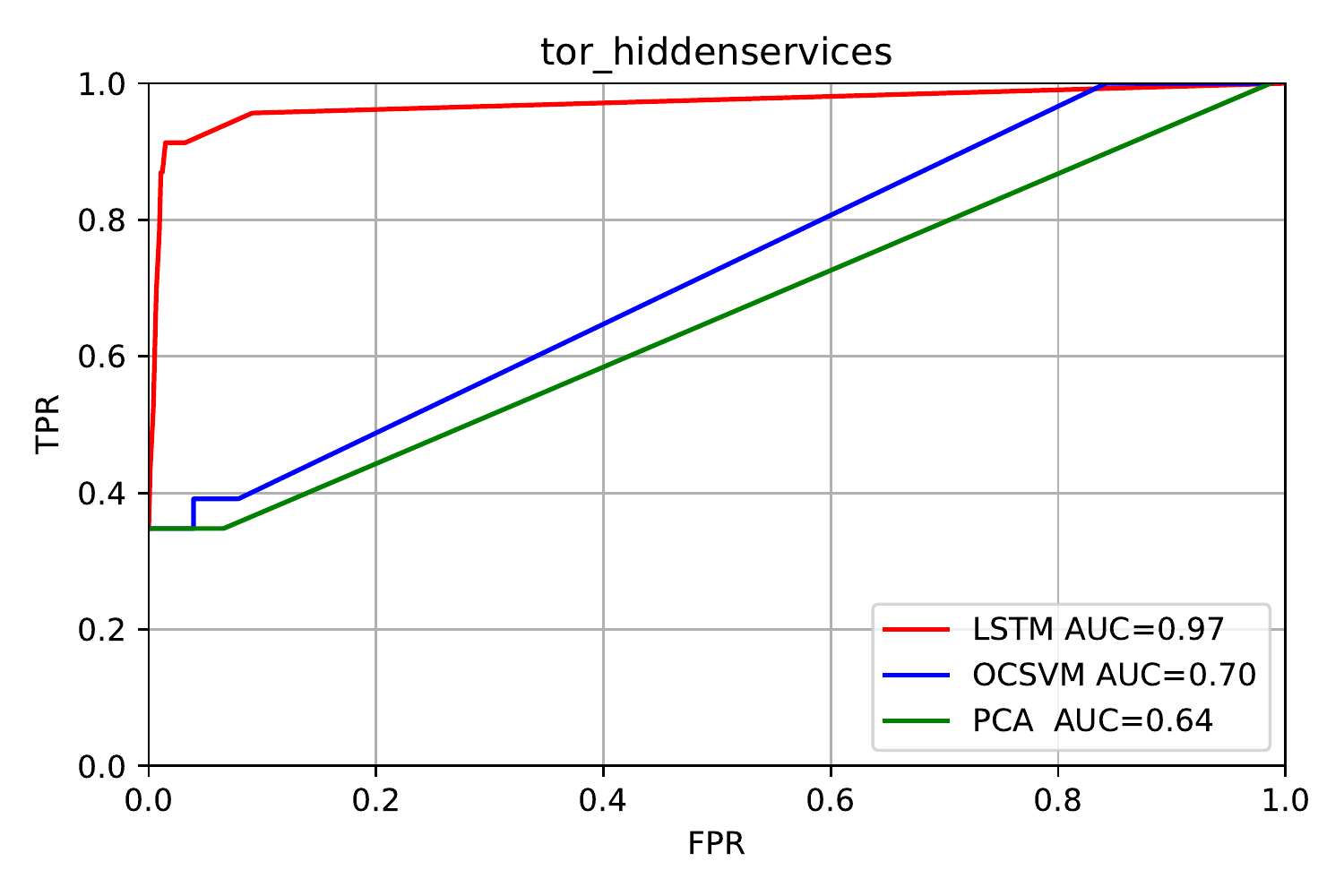}
\end{subfigure}
\caption{ROC curves for attack scripts for Struts CVE-2017-5638}
\label{fig:strutsrocs}
\end{figure*}

In Figure~\ref{fig:strutsrocs} we show the ROC curves for PCA, OCSVM, and LSTM for all 15 attack scripts for the Struts application with the Equifax exploit (CVE-2017-5638). We generated similar ROC curves for all applications, exploits, and attack scripts, but omit them here due to space limitations. 

